\documentclass{aa}  

\usepackage{color}
\usepackage{subfigure}
\usepackage{graphicx}
\usepackage{epsfig}
\usepackage{hyperref}
\usepackage{float}
\usepackage{multirow}
\newcommand{\grad}{$^{\circ}$~}

\def\xmm{XMM-{\it Newton}}
\newfont{\gwpfont}{cmssq8 scaled 1000}
\newcommand{\rexcess}{{\gwpfont REXCESS}}

\usepackage{txfonts}
\begin{document}

   \title{ATCA observations of the MACS-Planck Radio Halo Cluster Project}
   \titlerunning{Radio observations of an intermediate redshift cluster sample}
   \subtitle{II. Radio observations of an intermediate redshift cluster sample}
   \authorrunning{Martinez Aviles et al. 2}
   \author{G. Martinez Aviles, \inst{1} \fnmsep\thanks{gerardo.martinez.aviles@gmail.com}
          M. Johnston-Hollitt \inst{2,3},         
                  C. Ferrari\inst{1},
          T. Venturi \inst{4},                            
           J. Democles \inst{5},          
           D. Dallacasa \inst{6,4},
          R. Cassano \inst{4},
          G. Brunetti \inst{4},
          S. Giacintucci \inst{7},
          G. W. Pratt \inst{5},
          M. Arnaud \inst{5},
          N. Aghanim \inst{8},
          S. Brown \inst{9},
          M. Douspis \inst{8},
         G. Hurier \inst{8},
          H. T. Intema \inst{10},
          M. Langer \inst{8},  
          G. Macario\inst{1},
          \and
          E. Pointecouteau\inst{11,12}
          }

   \institute{Universit\'e C\^ote d'Azur, Observatoire de la C\^ote d'Azur, CNRS, Laboratoire Lagrange, Bd de l'Observatoire, CS 34229, 06304 Nice cedex 4, France 
         \and
             School of Chemical \& Physical Sciences, Victoria University of Wellington, PO Box 600, Wellington 6140, New Zealand
                 \and
Peripety Scientific Ltd., PO Box 11355 Manners Street, Wellington 6142, New Zealand
         \and
                 INAF- Istituto di Radioastronomia, via P. Gobetti 101, I-40129, Bologna, Italy
                 \and
Laboratoire AIM, IRFU/Département d'Astrophysique -- CEA/DRF -- CNRS -- Université Paris Diderot, Bât. 709, CEA-Saclay, 91191 Gif-sur-Yvette Cedex, France           
                 \and
                 Department of Physics and Astronomy, UniBO, Via Gobetti 93/2, I-40129 Bologna, Italy
                 \and
Naval Research Laboratory, 4555 Overlook Avenue SW, Code 7213, Washington, DC 20375, USA   
         \and     
Institut d'Astrophysique Spatiale, CNRS, Univ. Paris-Sud, Universit\'e Paris-Saclay, B\^at. 121, 91405 Orsay cedex, France
                 \and
Department of Physics $\&$ Astronomy, University of Iowa
         \and
Leiden Observatory, Leiden University, Niels Bohrweg 2, NL-2333CA, Leiden, The Netherlands
                 \and
CNRS; IRAP; 9 Av. colonel Roche, BP 44346, F-31028 Toulouse cedex 4, France
         \and
        Université de Toulouse; UPS-OMP; IRAP; Toulouse, France         
         }
   \date{Received July 7, 2017; accepted September 6, 2017}
 
  \abstract
   {}
   {A fraction of galaxy clusters host diffuse radio sources whose origins are investigated through multi-wavelength studies of cluster samples. We investigate the presence of  diffuse radio emission in a sample of seven galaxy clusters  in the largely unexplored intermediate redshift range (0.3 < $z$ < 0.44). 
    }
   {In search of diffuse emission, deep radio imaging of the clusters are presented from  wide band (1.1-3.1 GHz), full resolution ($\sim$5 arcsec)  observations  with the Australia Telescope Compact Array (ATCA). The visibilities were also imaged at lower resolution after point source modelling and subtraction and after a taper was applied to achieve better sensitivity to low surface brightness diffuse radio emission. In case of non-detection of diffuse sources, we set upper limits for the radio power of injected diffuse radio sources in the field of our observations. Furthermore, we discuss the dynamical state of the observed clusters based on an X-ray morphological analysis with \xmm.}
   {We detect a giant radio halo in PSZ2 G284.97-23.69 (z=0.39) and a possible diffuse source in the nearly relaxed cluster PSZ2 G262.73-40.92 (z=0.421). Our sample contains three highly disturbed massive clusters without clear traces of diffuse emission at the observed frequencies. We were able to inject modelled radio halos with low values of total flux density to set upper detection limits; however, with our high-frequency observations we cannot exclude the presence of RH in these systems because of the sensitivity of our observations in combination with the high z of the observed clusters. 
    }
   {}


   \maketitle
%

\section{Introduction}\label{sect:INTRO}

   \begin{table*}[]
      \caption[]{Information concerning the clusters of the MACS-Planck RHCP ATCA sample. The names and alternative names of the clusters appear in Cols. 1 and 2. The values RA and DEC correspond to the coordinates for centring the ATCA observations (Cols. 3 and 4). The Planck  mass, X-ray luminosity, and redshift (Cols. 5, 6, and 7, respectively) are taken from \cite{2016A&A...594A..27P}. $\star$ See Sect. \ref{sect:SELECTION} for a deeper discussion on PLCK G334.8-38.0.}
         \label{tab:sample}
     $$ 
         \resizebox{\width}{!}{\begin{tabular}{ccccccc}
            \hline
            \hline
            \noalign{\smallskip}
Cluster name &   Alternative name(s) &   RA      &   DEC & M$_{500}$ (SZ)   &  L$_{500}$   & Redshift  \\
& & (h m s)   &  (\grad $^{\prime}$ $^{\prime\prime}$)     & ($\times$ 10$^{14}$ M$_{\odot}$) &  ($\times$ 10$^{44}$ erg/s)    &     \\
            \noalign{\smallskip}
            \hline
            \hline
            \noalign{\smallskip}
PSZ2 G285.63-17.23 & PSZ1 G285.62-17.23 &  08 43 44.40 & -71 13 14.00 & 6.64$\pm$0.40 & 4.45$\pm$0.08 & 0.35\\
                   & PLCK G285.6-17.2   &              &              &               &        & \\
                    \noalign{\smallskip}
            \hline  
             \noalign{\smallskip}               
PSZ2 G262.73-40.92  & PSZ1 G262.72-40.92 &  04 38 17.20 & -54 19 25.10 & 7.46$\pm$0.36 & 9.94$\pm$0.47 & 0.421\\
                    & SPT-CLJ0438-5419  &                &               &              & & \\
                    & ACT-CL J0438-5419 &                &               &              & & \\
                    & PLCK G262.7-40.9  &                &               &              & & \\
             \noalign{\smallskip}
            \hline        
             \noalign{\smallskip} 
PSZ2 G277.76-51.74 & PSZ1 G277.75-51.71 & 02 54 16.70 & -58 56 44.00 & 9.69$\pm$0.38 & 9.46$\pm$0.07 & 0.438\\
                   & SPT-CLJ0254-5857 &               &              &               &            & \\
                   & PLCK G277.8-51.7 &               &              &               &            & \\  
             \noalign{\smallskip}
            \hline      
             \noalign{\smallskip}
PSZ2 G286.28-38.36 & PSZ1 G286.27-38.39 & 03 59 10.20 & -72 04 46.10 & 5.94$\pm$0.40 & 4.07$\pm$0.02 & 0.307\\
                   & PLCK G286.3-38.4   &             &              &               &          & \\
             \noalign{\smallskip}
            \hline  
             \noalign{\smallskip} 
PSZ2 G271.18-30.95 &  PSZ1 G271.18-30.95 & 05 49 19.50 & -62 05 16.00 & 7.37$\pm$0.32 & 18.95$\pm$0.16 & 0.37 \\
                   &  SPT-CLJ0549-6205   &             &              &               &     & \\              
                   &  PLCK G271.2-31.0   &             &              &               &     & \\ 
             \noalign{\smallskip}
            \hline  
             \noalign{\smallskip} 
PSZ2 G284.97-23.69 & PLCKESZ G284.99-23.70 & 07 23 18.40 & -73 27 20.60 & 8.39$\pm$0.34 & 16.91$\pm$0.27 & 0.39 \\
                   &  PLCK G285.0-23.7    &             &             &               &    &  \\    
                        \noalign{\smallskip}
                        \hline
                        \noalign{\smallskip}            
PLCK G334.8-38.0$\star$ & - & 20 52 16.80 & -61 12 29.40  & - & - & 0.35\\                        
             \noalign{\smallskip}
            \hline 
            \hline                       
\end{tabular}}
     $$ 
   \end{table*}

Observations of diffuse synchrotron radio emission in clusters of galaxies provide evidence of the interactions between ultra-relativistic particles and magnetic fields in the intracluster medium (ICM). At present, there are three main classes of large-scale diffuse radio emission in clusters: relics, haloes, and mini-haloes.

Radio relics are large ($\sim$1 Mpc) polarized diffuse sources generally elongated in shape that are present in the peripheral regions surrounding galaxy clusters. The mechanism responsible for their creation is associated with shocks generated in the ICM caused by a merger \citep{2014Brunetti}, although recently some relic-radio galaxy association has been discovered \citep[see e.g.][]{2017NatAs...1E...5V}. Radio mini-haloes are much smaller ($\sim$100 - 500 kpc) roundish sources present in cool-core clusters \citep{2017ApJ...841...71G} that surround a radio-loud active galactic nucleus (AGN) present in the brightest cluster galaxy (BCG) \citep{2009A&A...499..371G}. The origin of mini-halos is explained by a population of relativistic particles re-accelerated in the turbulence generated in the ICM by the mechanically powerful AGN or by gas sloshing in the cluster cool core \citep{2008ApJ...675L...9M}, although these explanations are still debated.

Finally, radio halos (RHs) are Mpc-scale, low-surface brightness radio sources observed to be centred in galaxy clusters with similar morphologies as the X-ray emission. Giant RH are the focus of this paper. In recent years, the knowledge of the origin of RHs has moved towards a general consensus. The generally accepted scenario for the mechanism responsible for this kind of diffuse emission is the re-acceleration of relativistic electrons by the large-scale turbulence generated in cluster mergers \citep[see e.g.][]{2001Brunetti, 2001ApJ...557..560P, 2014Brunetti}. In agreement with this model, it is known that RHs tend to occur more frequently in massive galaxy clusters \citep[see][for a recent discussion]{2015A&A...580A..97C}, and most of these show evidence of being merging systems. Moreover, a series of scaling relations between the thermal and non-thermal properties of galaxy clusters have also been found; i.e. P$_{1.4}$-L$_{x}$, P$_{1.4}$-Mass, P$_{1.4}$-Y$_{500}$, where P$_{1.4}$ and Y$_{500}$ are the radio power of haloes at 1.4 GHz and the cluster integrated Sunyaev-Zel'dovich (SZ) signal within R$_{500}$\footnote{R$_{500}$ is the radius at which the mean mass density is 500 times the critical density at the cluster redshift.} \citep[see e.g][]{2012MNRAS.421L.112B, 2013ApJ...777..141C}. The connection between mergers and the presence of a RH depends on a complex combination of mechanisms and energy budgets. There are a few cases reported in the literature of the presence of RHs in cool-core clusters or of clusters being minor mergers \citep[see e.g.][]{2016MNRAS.459.2940K, 2014MNRAS.444L..44B, 2015MNRAS.454.3391B, 2017MNRAS.466..996S}, and there is also evidence of merging clusters without detected RHs \citep{2010Cassano, 2011Russell}.

Fundamental questions about the micro-physics of electron acceleration and transport mechanisms still need to be answered to characterize the non-thermal physics of clusters. Statistical studies of RHs and their host systems can shed light on the mechanisms needed to complete the picture of the physics of RHs and their connection with cluster mergers. Current knowledge of RHs is mostly based on high mass galaxy cluster samples. An important study on RH occurrence is the Giant Metrewave Radio Telescope (GMRT) Radio Halo Survey (GRHS) \citep{2007A&A...463..937V, 2008Venturi} and the extended sample (E-GRHS) \citep{2013A&A...557A..99K, 2015A&A...579A..92K} with the aim of exploring the origin and occurrence of RHs and their connection with the dynamical state of the host systems. The authors picked galaxy clusters with a redshift range z = 0.2-0.4, X-ray luminosities L$_{x}$ (0.1-2.4 keV) > 5 $\times$ 10$^{44}$ erg s$^{-1}$ and declinations $\delta$ > -31\grad to ensure a good uv coverage with the GMRT. This sample is however effectively limited to z < 0.33. Mass-based selections are motivated by the assumption that the selection of high X-ray luminosity merging systems have a greater chance of detecting diffuse radio sources. To start filling this observational gaps, previous works investigated the presence of diffuse radio emission with an unbiased sample on X-ray morphology and in a wide range of masses \citep{2016MNRAS.459.2525S}.

Although a number RHs have been discovered at z > 0.3, a complete statistical sample is missing in this redshift regime \citep[see][]{2012A&ARv..20...54F}. On the other hand, it is expected from models \citep[see][]{2004JKAS...37..589C, 2006Cassano} that a larger number of RH occurrence may appear at z=0.3-0.4. This is simply because most of the energy budget in the hierarchical growth of clusters is dissipated via massive mergers in this redshift range. In this paper, we present the results of a series of radio observations of seven galaxy clusters in this redshift regime performed with the ATCA telescope, together with an X-ray analysis of the dynamical status of the target clusters. 

This work is organized in the following way: Sect. \ref{sect:SELECTION} describes the criteria for the sample selection. In Sect. \ref{sect:REDUCTION} we present the observations together with the data reduction and image reconstruction strategy. The analysis of the high resolution, tapered, and compact source subtracted radio images appears in Sect. \ref{sect:radioimages}. In Sect. \ref{sect:dynamic} we present the X-ray morphological analysis of the cluster sample. The results of the paper appear in Sect. \ref{sect:results}. Finally, the discussion and conclusions are presented in Sect. \ref{sect:conclusions}. Throughout this paper, we adopt the  $\Lambda$CDM cosmology with the values $H_{o}$=70 km s$^{-1}$ Mpc$^{-1}$, $\Omega_{M}$=0.3, $\Omega_{\Lambda}$=0.7.

   \begin{table*}
      \caption[]{Details of the observations towards the clusters selected for observations with ATCA. Cluster name (Col. 1); dates of observations (Col. 2) with different array configurations (Col. 3); observation time (Col. 4); and phase calibrator (Col. 5). The central frequency is at 2.1 GHz and the total observed bandwidth is 2 GHz.}
         \label{tab:obs}
     $$ 
         \begin{tabular}{cccccc}
            \hline
            \hline
            \noalign{\smallskip}
            Cluster name &      Date      &   Config.   & Observation   & Calibrator \\
                  &                &             &  time (min.) & \\
            \noalign{\smallskip}
            \hline
            \hline
            \noalign{\smallskip}
PSZ2 G285.63-17.23 &  2015-Jun-12  & 1.5C &  549 & PKS B0606-795 \\                                
                         &  2012-Jun-29  & 750A &  442 & PKS B0606-795 \\
                         &  2012-Jun-07-08  & 6D & 1132  & PKS B0637-752   \\
                    \noalign{\smallskip}
            \hline  
             \noalign{\smallskip}               
   PSZ2 G262.73-40.92   &  2013-Mar-05  & 6A   &  353 & PKS B0420-625 \\
                    &   2013-Feb-02  & 750C &  354 & PKS B0420-625 \\
                    &  2012-Nov-23   & 1.5C &  354 & PKS B0420-625 \\
             \noalign{\smallskip}
            \hline        
             \noalign{\smallskip} 
   PSZ2 G277.76-51.74  &  2013-Sep-03-04   & 1.5A  &  1140  & PKS B0302-623\\
                                         &  2013-Jul-31      & 750D  &  1061  &PKS B0302-623 \\
                                         &  2013-Aug-01      & 750D  &  152  & PKS B0302-623\\
                                         &  2013-May-11-12           & 6C &  1188  & PKS B0302-623 \\
             \noalign{\smallskip}
            \hline      
             \noalign{\smallskip}
      PSZ2 G286.28-38.36  & 2013-Mar-05   & 6A   &  384 & PKS B0252-712\\
                    &   2012-Feb-02   & 750C &  354 & PKS B0252-712\\           
                    &  2012-Nov-23    & 1.5C &  413 & PKS B0252-712\\
             \noalign{\smallskip}
            \hline  
             \noalign{\smallskip} 
         PSZ2 G271.18-30.95 &  2013-Mar-05  & 6A &  354 & PKS B0420-625\\                                                                  
                     &  2013-Feb-02  & 750C & 294 & PKS B0420-625\\
                     & 2012-Nov-23  & 1.5C & 353 & PKS B0420-625\\
             \noalign{\smallskip}
            \hline  
             \noalign{\smallskip} 
     PSZ2 G284.97-23.69     & 2012-Jun-08 & 6D & 704 & PKS B1036-697 \\
                          & 2012-Jun-09 & 6D & 523 & PKS B0606-795 \\
                          & 2012-Jun-29 & 750A & 531 & PKS B0606-795 \\
                          & 2013-Sep-06 & 1.5A & 804 &  PKS B0606-795  \\
             \noalign{\smallskip}
            \hline
             \noalign{\smallskip}                      
      PLCK G334.8-38.0 &  2013-Sep-03-04  & 1.5A  &  1131 & PKS B0302-623\\                        
               &  2013-Jul-30          & 750D  &  1203 & PKS B1934-638\\
               &  2013-May-12          & 6C    &  1297 &  PKS B1934-638\\
             \noalign{\smallskip}
            \hline 
            \hline                       
\end{tabular}
     $$ 
   \end{table*}

\section{Sample selection}\label{sect:SELECTION}   

This paper presents the results of a subsample of the MACS-Planck Radio Halo Cluster Project (RHCP). The project was conceived as a continuation of the E-GRHS project (see Sect. \ref{sect:INTRO}) by extending the statistics of RHs up to z=0.45. The Macs-Planck RHCP consists of a total 48 galaxy clusters, all of which were taken from the MACS Brightest Cluster X-ray catalogue \citep{2010MNRAS.407...83E} and the list of newly discovered Planck SZ clusters confirmed with the \xmm\  validation programme
\citep[available at the time of the proposal; see][]{2011A&A...536A...9P}. For the project, all the clusters from both samples that are located in the range z = 0.3-0.45 were selected. In our sample we have a total of 33 MACS clusters and 15 Planck SZ clusters. The scope of this paper is to publish the analysis of the ATCA subsample. Information about the complete sample will appear in a forthcoming paper (Venturi et al., in prep.).

From the total number of selected clusters, 32 were lacking published radio information. The declination range of the sample required observations to be carried out with two different telescopes, depending on the declination of the targets. Based on the visibility of the targets and the uv-coverage constraints to achieve the desired sensitivities to detect diffuse radio emission, a threshold of $\delta$ = -40\grad was defined. This threshold divided the total sample into two groups:

\vspace{2mm}
a) Those lying in the range of $\delta$ > -40\grad (25 clusters in total) were observed with the GMRT at $\sim$ 325 MHz, and the results of these observations will appear in a forthcoming paper (Venturi et al., in prep.). 

\vspace{2mm}
b) The remaining seven galaxy clusters ATCA observations are centred at 2.1 GHz with a bandwidth of 2000 MHz. Information about the clusters from the ATCA sample appears in Table \ref{tab:sample}. It is worth mentioning that it was a matter of pure chance that all the clusters observed with the ATCA were Planck clusters.

It is important to mention that the measurements of redshift and mass for PLCK G334.8-38.0 are very challenging because they are a low-mass triple system. Having very poor X-ray statistics we cannot analyse the system by proceeding in the same way we have for the other clusters of the sample. However, the mass of the components, even taken all together (M $\sim$ 3.4 $\times$ 10$^{14}$M$_{\odot}$), is significantly lower than the typical mass of known radio loud systems. 

\begin{figure*}[]
   \includegraphics[width=\hsize]{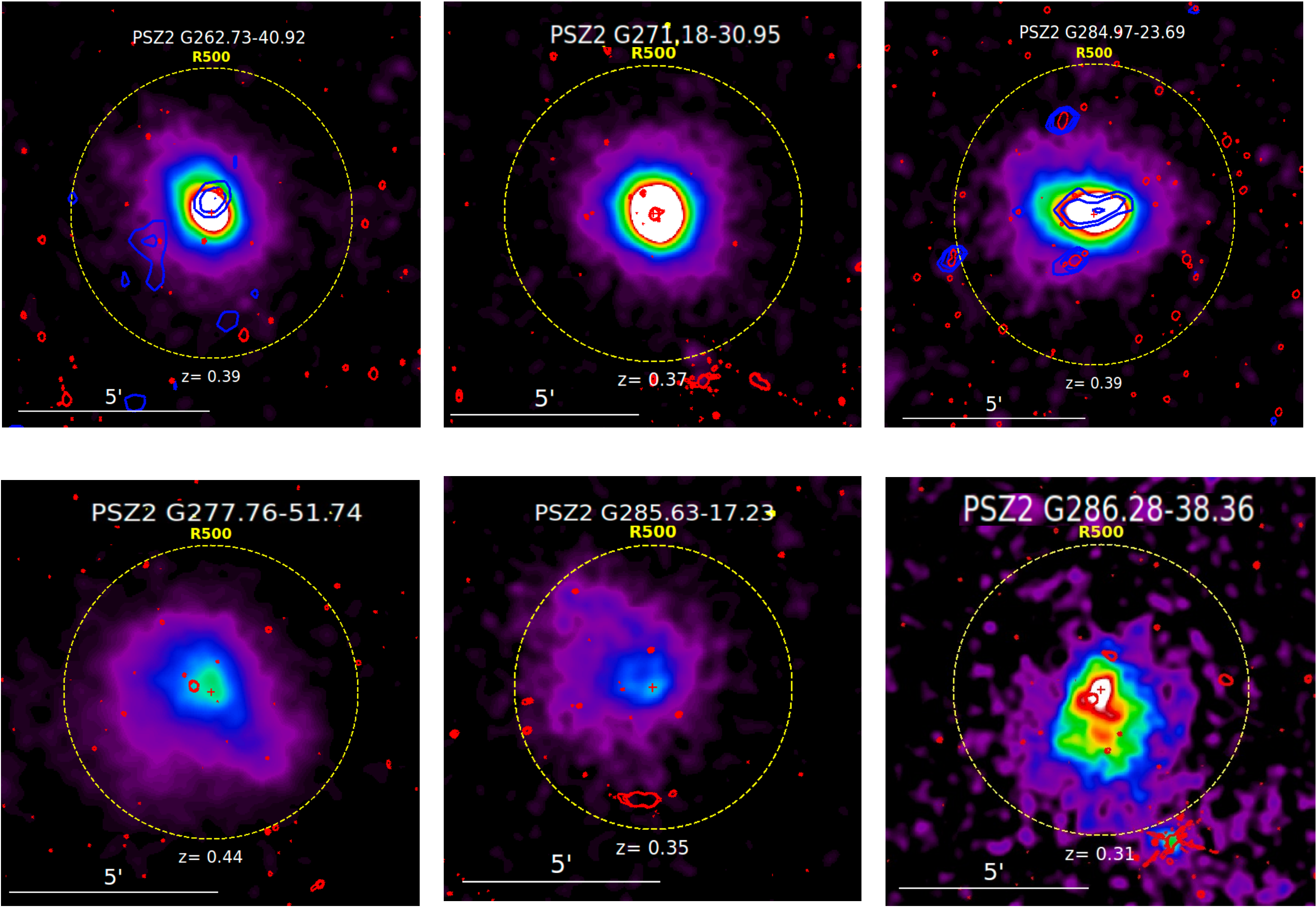}
      \caption{\xmm\ smoothed images in the [0.3-2] keV band of the ATCA cluster sample overlaid with the 3$\sigma$ $\times$ (1, $\sqrt{2}$, 2) contours of the high resolution wide band radio maps (see Table \ref{table:maps} and Figs. \ref{fig:Figwide1} and \ref{fig:Figwide2}) in red and the same contour levels for the Block 3 low resolution compact source subtracted maps for the two clusters with diffuse radio emission (see Sect. \ref{sect:results}) in blue. The background of the X-ray images has been subtracted. The X-ray images are corrected from surface brightness dimming with redshift, divided by emissivity in the energy band, taking in account absorption from the Galaxy and the response of the instrument. Finally the images are scaled to a self-similar model. The colours are selected so that the images would look identical when scaled by their mass if they were at the same distance; this gives us a visual hint of the dynamical state of the clusters.}
        \label{fig:Xrayoverlay}
\end{figure*}

\section{Radio observations and data reduction}\label{sect:REDUCTION}

   Radio observations of the ATCA sample were carried out with three separate array configurations for each observed cluster (Table \ref{tab:obs}) using the Compact Array Broadband Backend (CABB) correlator with a central frequency of 2.1 GHz and spanning 1.1-3.1 GHz (project ID C2679). Observations were carried out in continuum mode with the correlator set to produce 2000 $\times$ 1 MHz channels. Details of the observations can be found in Table \ref{tab:obs}. The primary flux scale was set relative to the unresolved source PKS\,B1934-638 for which the detailed spectral behaviour is well understood. The amplitude gain variations were checked during the calibration of each observation for each sub-band image such that these variations were not higher than $\sim$2$\%$.\footnote{see http://www.atnf.csiro.au/observers/memos/d96783~1.pdf.}
   
Radio frequency interference (RFI) and bad channels were excised manually from primary and secondary calibrators, as well as the target, using a combination of clipping algorithms and visual inspection via the MIRIAD task PGFLAG \citep{1995ASPC...77..433S}. It was necessary to perform calibration on narrower frequency intervals owing to the nature of CABB data. After a number of trials, we determined that four sub-bands of $\sim$500 MHz produce the optimal results for these data. Thus, the target, primary, and secondary calibrator data sets were divided into the required sub-bands coined Block 1 (from 2.631 GHz to 3.100 GHz), Block 2 (from 2.131 GHz to 2.630 GHz), Block 3  (from 1.631 GHz to 2.130 GHz), and Block 4 (from 1.130 GHz to 1.630 GHz). Each sub-band was then self-calibrated in MIRIAD via the task SELFCAL. Finally, the self-calibration solutions were saved with the task UVAVER.

\section{Radio analysis}\label{sect:radioimages}

\subsection{High resolution images}\label{subsect:hiresimag}

Four images were created, one for each of the four sub-bands, out to the primary beam via the task INVERT, and a Steer CLEAN \citep{1984A&A...137..159S} was applied to all sources within the primary beam. The sub-band images were then convolved to a common size by applying a Gaussian corresponding to the lowest resolution image and were then added together to create a final wide-band image (see Table \ref{table:maps}). The noise levels in the sub-band images were very similar for the three highest bands and only differed slightly in the lowest frequency band; it was not necessary to weight the images relative to the sensitivities in the mosaicking process. 

The radio contours of the central Mpc diameter area from our high resolution images are overlaid on the \xmm\ X-ray images in Fig. \ref{fig:Xrayoverlay} (yellow contours). In Figs. \ref{fig:Figwide1} and \ref{fig:Figwide2} we show the high resolution ATCA radio maps for the wide field and zoomed cluster regions\footnote{The high resolution images from PSZ2 G284.97-23.69 have been published in \citet{2016A&A...595A.116M}.}. The final deep ATCA images root mean squared (rms) noise is measured with AIPS TVSTAT at the field centre in regions inside the primary beam without any trace of point sources or diffuse emission. The information of the images (rms noises and resolutions) is shown in Table \ref{table:maps}. A visual inspection of the seven high resolution images does not reveal obvious presence of diffuse radio emission in the central area in which the clusters of the sample lie.

\begin{table*}
 \centering
      \caption[]{Properties of the full resolution and Block 3 compact source subtracted tapered 2000 MHz band-width radio maps centred at 2.1 GHz. The rms noise of the images appears in Col. 2, while the resolutions and position angles are shown in Cols. 3 and 4, respectively.}\label{table:maps}
  \begin{tabular}{|c|c|c|c|c|c|c|}
    \hline
    \multirow{2}{*}{Cluster name} &
      \multicolumn{2}{c|}{rms noise ($\mu$Jy/beam)} &
      \multicolumn{2}{c|}{Beam size ($^{\prime\prime}$ $\times$ $^{\prime\prime}$)} &
      \multicolumn{2}{c|}{PA (deg)} \\
    & Full resolution  & Taper & Full resolution & Taper & Full resolution & Taper\\
    \hline
    PSZ2 G285.63-17.23 & 22.4 & 92 & 6.85 $\times$ 4.07  & 43 $\times$ 17 & -8.13  & 89.07 \\
    \hline
    PSZ2 G262.73-40.92  & 19.7 & 83 & 6.39 $\times$ 4.15 & 32 $\times$ 25  & -2.45  &  -7.67 \\
    \hline
    PSZ2 G277.76-51.74 & 15.4 & 65  & 5.55 $\times$ 4.25 & 35 $\times$ 28 & 1.99 &  58.04 \\
    \hline
    PSZ2 G286.28-38.36 & 19.4 & 57 & 5.00 $\times$ 4.46  & 28 $\times$ 26 & -11.00 &  -46.47 \\
    \hline
    PSZ2 G271.18-30.95  & 22.1  & 74 & 5.80 $\times$ 4.21 & 31 $\times$ 25 & -1.30  &  -28.70 \\
    \hline
    PSZ2 G284.97-23.69 & 11.3  & 51  &  5.20 $\times$ 4.38 &  31 $\times$ 23 & -42.31 & -81.17  \\
    \hline
    PLCK G334.8-38.0 & 16.1 & 71  & 3.19 $\times$ 2.20 & 34 $\times$ 23 & 9.96 &  2.96 \\
    \hline
  \end{tabular}
\end{table*}

\subsection{Tapering and diffuse source search}\label{subsect:taper}

We proceeded to subtract the compact radio sources from the uv data. Using the self-calibrated data, we imaged the visibilities for the longest baselines (uv range from 3.6 to 40 k$\lambda$, which translates to angular scales from $\sim$5 to $\sim$57 arcseconds) and subtracted the corresponding clean components from the self-calibrated uv data via the MIRAD task UVMODEL. This creates a new, re-imaged, uv data set. Again, none of the seven clusters showed the obvious presence of diffuse emission in the central area in the compact source subtracted high resolution images.

In order to investigate the possible presence of low-surface brightness radio emission in the clusters, the point source subtracted visibilities were re-imaged with a robust=0.5 weighting and by applying a $\sim$ 9 k$\lambda$ Gaussian taper. Traditionally, measurements of flux densities of RHs are reported at the central frequency of 1.4 GHz. It is worth mentioning that from the typical spectral behaviour of radio halos $(\alpha$ $\gtrsim$ 1 synchrotron spectral index\footnote{In this paper we use the convention S($\nu$) $\propto\nu^{-\alpha}$, where  S($\nu$) is the radio flux density.}), lower luminosities are expected for higher frequencies. Because of its lower central frequency, the Block 4 sub-band would be ideal to do flux density measurements, but it is also strongly affected by RFI (50-60 $\%$ of the data was flagged). In our images we have comparable rms sensitivities for the frequency range running from Block 3 to Block 1. This set of conditions makes Block 3 centred at $\sim$ 1.9 GHz the most suitable sub-band in which to carry out the flux density measurements. In Table \ref{table:maps} we report the main features for the set of the source subtracted Block 3 tapered images. 

In search of residual diffuse emission in the cluster areas we set the classical 3$\sigma$ contours in the compact source subtracted Block 3 tapered images. In some cases, we found contours that correspond to positions of point sources visible in the high resolution maps due to an imperfect point source subtraction or to the presence of some extended emission candidates (see Sect. \ref{sect:results}). Radio images that showed no evidence of the presence of diffuse sources were used to set detection limits for possible diffuse radio emission, as described below.

\subsection{Upper limit determination}\label{subsect:injections}
  
To set the detection upper limits for possible diffuse radio emission in our ATCA observations, we took a similar approach to that used in, for example, \cite{2008Venturi}. Specifically, our procedure consists of the following steps:

\vspace{1mm}
1) We simulated a diffuse source with the MIRIAD task IMGEN. The model consist of five low surface brightness concentric disks, of which the biggest has a diameter of 1 Mpc in the corresponding redshift of the respective galaxy cluster image. To reproduce the typical profile of a RH, these disks area percentages are from largest to smallest:100$\%$, 60$\%$, 33$\%$, 25$\%,$ and 12$\%$. These disks contribute, respectively, the following percentage of the total flux of the simulated source: 72$\%$, 20$\%$, 5$\%$, 2$\%,$ and 1$\%$.

\vspace{1mm}
2) The simulated source was added into the compact source subtracted uv data of the corresponding observation on the sub-set coined Block 3 ($\sim$500 MHz wide, see Sect. \ref{sect:REDUCTION}) for each galaxy cluster. This step was performed with the MIRIAD task UVMODEL, which generates a new uv data set of the real data plus the simulated source. To perform the injection of the spheric models we picked the 1 Mpc diameter circle centred on the cluster coordinates and five 1 Mpc areas inside the primary beam, without any trace of point sources and where there is no appearance of 3$\sigma$ contours due to artefacts on the original image plane.

\vspace{1mm}
3) The model injected uv data were imaged at low resolution applying a taper with the following parameters on the MIRIAD task invert: FWHM=20, robust=0.5, and cell=4; this allowed us to achieve more sensitivity to the kind of emission of the simulated halos, as in Sect. \ref{subsect:taper}. 

\vspace{1mm}
4) The injected fluxes of the different modelled disks were increased until we noticed that 3$\sigma$ contours appeared uniformly on the injection areas with a size large enough ($\sim$500 kpc diameter) to allow us to recognize it as a candidate RH. 

\vspace{2mm}
\noindent We were able to recover from $\sim$50$\%$ to $\sim$70$\%$ of the original injected flux by measuring the total flux density in the areas of the image plane where the injections were carried out. In the range of rms noises and resolutions of our images (see Table \ref{table:maps}), we determined that RH models with a total flux density of 3-5 mJy are the lowest values that can be injected in our images to be considered as upper limits of detection. The radio powers for the detection limits calculated at 1.4 GHz with whole injected flux densities, assuming a spectral index of 1.3, are shown in Table \ref{table:upperlimits}.

\section{X-ray dynamical state of the ATCA clusters}\label{sect:dynamic}

All the clusters of the ATCA sample benefit from X-ray observations from the \xmm\ space telescope as part of the validation programme of Planck cluster candidates.
The X-ray data processing is detailed in \citet{2011A&A...536A...9P}. The  cluster mass, $M_{500}$, and corresponding $R_{500}$  are derived iteratively using the low scatter $M_{500}$--$Y_{\rm X}$ scaling relation from \citet{2011A&A...536A...9P}, where $Y_{\rm X}$ is the product of the gas mass within $R_{500}$ and the X-ray temperature in the $[0.15-0.75]\,R_{500}$ aperture.  The  density profiles were derived from the surface brightness profile centred on the X--ray emission peak, using the PSF-deconvolution and deprojection method of \citet{2008A&A...487..431C}. From this analysis, we compute two morphological parameters:

\vspace{3mm}
\noindent a) The surface brightness concentration parameter $ C$,

\begin{equation}
\mathrm{C} = \frac {  S_{\rm X} (< R_{\rm in}) }   {S_{\rm X} (< R_{\rm out})}
,\end{equation}

\noindent
the ratio of  the surface brightnesses $S_{\rm X}$ within an inner aperture  $R_{\rm in}$ and a global aperture $R_{\rm out}$. The value $S_{\rm X}$
 is  the PSF-corrected surface brightness,  which is derived from the emission measure profile.   Introduced in \citet{2008A&A...483...35S} using $R_{\rm in} = 40$ kpc and 
$R_{\rm out}$ = 400 kpc, this parameter has been widely used to probe the core properties of clusters up to high redshift \citep[e.g.][]{2010A&A...521A..64S, 2010A&A...513A..37H, 2015MNRAS.447.3723P}.  Here we choose to use scaled apertures since the clusters of our sample cover a wide redshift range, with $R_{\rm in}= 0.1\times R_{500}$, corresponding to the typical size of the cool core, and $R_{\rm out} = 0.5 \times R_{500}$, which is a characteristic size for the total flux.

\vspace{3mm}
\noindent b) The X-ray centroid-shift w, as defined in, for example, \citet{2010A&A...514A..32B} within 10 circular apertures from 0.1 to 1 R$_{500}$ excising the first central aperture,

\begin{equation}
w = \left [ \frac{1}{N+1} \sum (\Delta_{i}-\left \langle \Delta \right \rangle )^{2} \right ]^{1/2} \times \frac{1}{R_{500}}
,\end{equation}

\noindent where $\Delta_i$ is the distance between the emission weighted centroid within the ith aperture and the X-ray peak and N is the number of apertures.

The centroid shift is computed on the background-subtracted, exposure-corrected co-added X-ray count images in the 0.3-2 keV energy band after removal and refilling of the point sources as in \citet{2010A&A...514A..32B}.

\begin{figure}[]
   \includegraphics[width=\hsize]{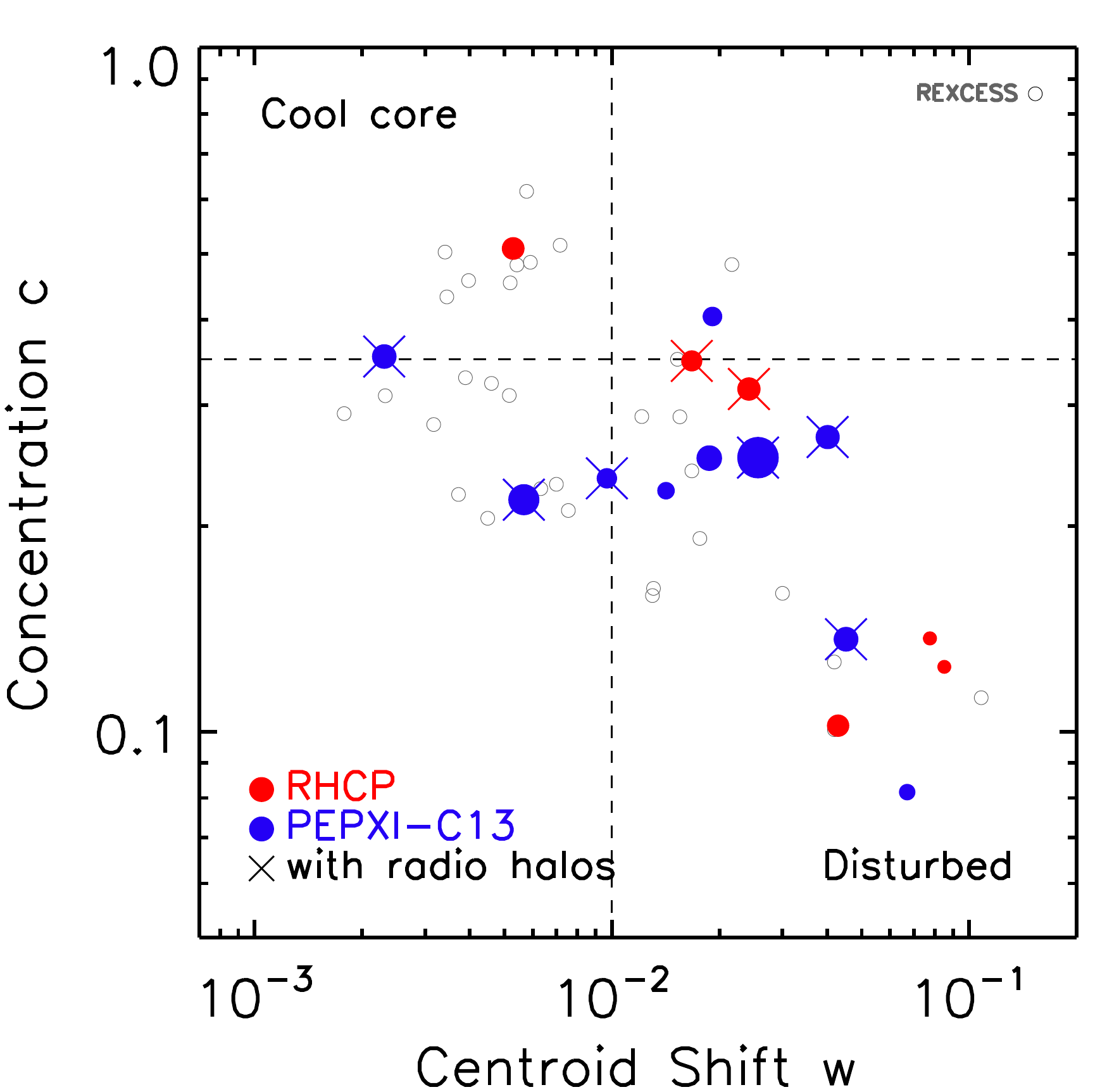}
      \caption{Concentration parameter ${C}$ vs. centroid shift $w$  for the galaxy clusters of the RHCP sample analysed in this paper (red filled circles).  For comparison, we also show the clusters in the sample of  \citet{2013ApJ...777..141C} that appear in the sub-sample of Planck clusters studied by \citet{2011A&A...536A..11P} (PEPXI-C13, blue  filled circles) for which we applied our algorithm to compute the w values within $R_{500}$.
The size of the circles is proportional to the $\log(M_{500})$, in the range   $[14.67$--$15.3]$, and clusters with detected radio haloes are indicated with a cross. The characteristic thresholds indicating  cool cores and morphologically disturbed systems (dashed lines) are from the \rexcess\  study \citep[]{2007A&A...469..363B}. The \rexcess\  clusters are shown as open circles.}
        \label{fig:Morpho}
\end{figure}

The results are shown in Fig. \ref{fig:Morpho}, where we overplot for comparison the positions of objects from the \rexcess\ \citep[open grey circles; see][for the centroid-shift values]{2010A&A...514A..32B}. Fig. \ref{fig:Morpho} also includes the systems in common between the samples of \citet{2013ApJ...777..141C} and  \citet{2011A&A...536A..11P} (blue points), for which we computed the concentration parameter and the centroid-shift values as described above for the ATCA sample.

The horizontal and vertical dashed lines of Fig. \ref{fig:Morpho} indicate characteristic threshold values of each parameter that are typically used to separate out cool-core and morphologically disturbed systems, respectively. Following \citet{2009A&A...498..361P}, we use a value of \textit{w} > 0.01 as indicative of a morphologically disturbed system. Similarly, we define targets with \textit{C} > 0.35, which is equivalent to the central density criterion used by \citet{2009A&A...498..361P}, as centrally peaked and thus cool-core systems.

We stress here that the limits indicated with dashed lines in Fig.\ref{fig:Morpho} were obtained for the local \rexcess\ sample. Also the sample analysed by \citep{2013ApJ...777..141C} is mostly comprised of lower redshift clusters compared to our targets. In addition, instead of using scaled apertures, \citet{2013ApJ...777..141C} adopted fixed physical sizes for computing the two parameters \textit{w} (500 kpc) and \textit{C} ($S_{x}$(<100 kpc)/$S_{x}$(500 kpc)). Their choice was based on the theoretical consideration that, for a typical $\sim$1 Mpc-size RH, 500 kpc is expected to delimitate the region in which the energy of the merger is dissipated in particle acceleration. These factors are however not expected to make our analysis significantly different from the approach of \citet{2013ApJ...777..141C}, since recent works have proven that the adopted morphological parameters neither depend significantly on the size of the central region selected to estimate them (e.g. good agreement between $w$ measured within $R_{500}$ and $0.5~R_{500}$) nor are limited by resolution issues up to z $\sim$1 clusters \citep[e.g.][Lovisari et al., private communication]{2017A&A...598A..61B}. Our new estimates of the \textit{C} and \textit{w} parameters, however, indicate that one of the radio-loud clusters in \citet{2013ApJ...777..141C} falls in the relaxed part of the plot.

\section{Results}\label{sect:results}

\subsection{Giant radio halo in PSZ2 G284.97-23.69}\label{subsect:Halo2850}
 
In \citet{2016A&A...595A.116M} we reported the discovery of a giant radio halo in PSZ2 G284.97-23.69 (named in our previous publication PLCK G285.0-23.7). Because PSZ2 G284.97-23.69 is the only confirmed giant RH discovered in this set of observations, it is important to discuss it in the complete context of the whole ATCA sample. PSZ2 G284.97-23.69 is the second most luminous and massive cluster in the sample (see Table \ref{tab:sample}). Based on its disturbed morphology (\textit{w} = 0.028 $\pm$ 0.001) and concentration parameter (C = 0.317 $\pm$ 0.005) it falls in a quadrant of dynamically disturbed systems, albeit with a relatively high \textit{C} parameter. The size and orientation of the 3$\sigma$ contours on the Block 3 tapered radio map have a remarkable coincidence with the X-ray morphology \citep[see Fig.4 in][]{2016A&A...595A.116M}. The detected RH seems to be slightly under-luminous compared to objects hosted by clusters in a similar mass range, as shown in the P$_{1.4 GHz}$  vs M$_{500}$ plot (purple star in Fig. \ref{fig:Fighalos}); although it is not a clear outlier like the RH discovered by \citet{2015MNRAS.454.3391B}. It is, however, one of the lowest luminosity radio halos detected at z > 0.35.

\subsection{Diffuse radio source in the cluster PSZ2 G262.73-40.92.}\label{subsect:source262} 

\begin{figure}[h]
   \includegraphics[width=10cm]{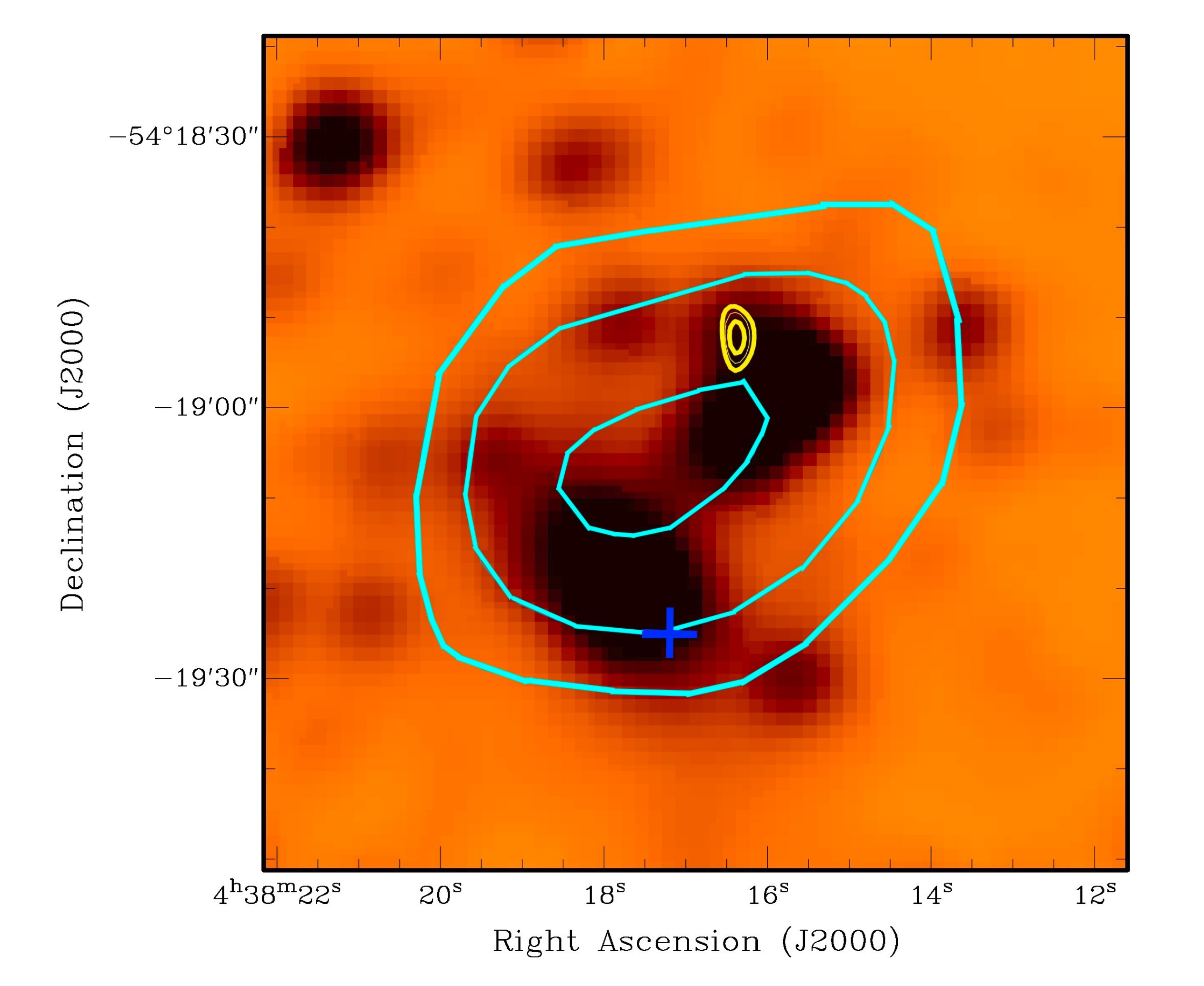}
      \caption{WISE 4.6 microns image of the central area of PSZ2 G262.73-40.92. The centre of the X-ray emission is indicated as a blue cross. Contours of the compact source, 3$\sigma$ $\times$ (1, $\sqrt{2}, 2$), subtracted tapered radio image appear in cyan, while yellow contours are the same contour levels from the high resolution Block 3 radio map.}
        \label{fig:262overlay}
\end{figure}

According to the X-ray morphological analysis, PSZ2 G262.73-40.92 is a merging, but probably moderately disturbed cluster, lying close to our cut line of cool-core clusters, with a concentration parameter C= 0.348 $\pm$ 0.005 and a centroid shift \textit{w} = 0.016 $\pm$ 0.001 (see Fig. \ref{fig:Morpho}). The tapered compact source subtracted image in the field of the cluster PSZ2 G262.73-40.92 shows the presence of a candidate diffuse source close to the cluster central position. The high resolution image reveals the presence of a very faint point source in the region of the diffuse emission, with a possible counterpart on the IR map (Fig. \ref{fig:262overlay}), for which we measured a flux density of 0.33 $\pm$ 0.04 mJy. We measured the total flux density inside the 3$\sigma$ contours (shown in cyan colour in Fig. \ref{fig:262overlay}) of the diffuse emission in the compact source subtracted tapered image in Block 3, which is characterized by a rms noise of 83 $\mu$Jy/beam (see Table \ref{table:maps}). We obtained a flux density of 1.18 $\pm$ 0.12 mJy inside the 3$\sigma$ contours. The region shows an elliptical shape, with a major axis of 63 arcseconds and a minor axis of 50 arcseconds, which for our cosmology corresponds to a physical size of 349 kpc $\times$  277 kpc at the cluster redshift. The diffuse source present in the residual map follows the emission from the galaxies in the WISE images (Fig. \ref{fig:262overlay}), and is offset from the cluster X-ray emission (Fig. \ref{fig:Xrayoverlay}, top, left panel), which makes the confirmation and possible classification of the source challenging with our data.

\subsection{Clusters without diffuse radio emission}\label{subsect:noradio}

 \subsubsection{Cool-core cluster PSZ2 G271.18-30.95}\label{subsect:Coolcore}

PSZ2 G271.18-30.95 has the highest X-ray luminosity of the ATCA sample, although it is not the most massive (see Table \ref{tab:sample}). Within the sample this cluster has also the highest concentration parameter (C = 0.509 $\pm$ 0.004) and the lowest centroid shift (\textit{w} = 0.005 $\pm$ 0.001). This positions PSZ2 G271.18-30.95 as a clear cool-core non-disturbed cluster. Unsurprisingly, as in $\sim$70$\%$ of cool-core clusters \citep{2015aska.confE..76G} our high resolution radio map shows the presence of a strong radio source coincident with the cluster centre (see Fig. \ref{fig:Xrayoverlay} top middle panel and Fig. \ref{fig:Figwide2} bottom panels) with a flux density of $\sim$ 10 mJy at $\sim$ 1.9 GHz. From the Hubble space telescope optical image there is an evident connection between the radio source and the BCG. Moreover, by measuring the flux densities of the central source in various sub-bands (Blocks 1 to 4, see Sect. \ref{sect:REDUCTION}) we estimate a spectral index for the central compact source of $\alpha$ $\approx$ 0.6. 

It is known that radio mini-halos are usually located in the centre of cool-core clusters, surrounding the central active radio galaxy and extending to the radius of the cluster cooling region \citep[see e.g.][and references therein]{2017ApJ...841...71G}. To further investigate the possible presence of a radio mini-halo, we did several tests by imaging at $\sim$ 10 arcsec resolution without compact source subtraction and using different values for the taper. None of our maps show evidence of extended diffuse emission at the typical mini-halo scales. We point out, however, that the presence of the compact source at the cluster centre and the presence of a strong source close to the cluster made the imaging and compact source subtraction problematic in this particular case.

\subsubsection{Highly disturbed clusters without radio emission}\label{subsubsect:disturbed}
Three of the seven galaxy clusters of the ATCA sample show evidence that they are both highly disturbed (see Fig. \ref{fig:Xrayoverlay} lower panels and the three red circles in the bottom right of Fig. \ref{fig:Morpho}) and have low concentration parameters. PSZ2 G277.76-51.74, the most massive cluster of the sample, also has the lowest concentration parameter (C = 0.102 $\pm$ 0.002) and shows a centroid shift \textit{w} = 0.042 $\pm$ 0.001. On the other hand, PSZ2 G285.63-17.23 has the highest centroid shift (\textit{w} = 0.081 $\pm$ 0.003, C = 0.124 $\pm$ 0.006). Finally PSZ2 G286.28-38.36 also shows signs of being highly disturbed (\textit{w} =  0.078  $\pm$ 0.005, C = 0.137 $\pm$ 0.008). None of these clusters show hints of diffuse radio emission on our maps.

\subsubsection{Detection limits}\label{subsubsectresults:limits}

\begin{figure*}
   \centering
   \includegraphics[width=\hsize]{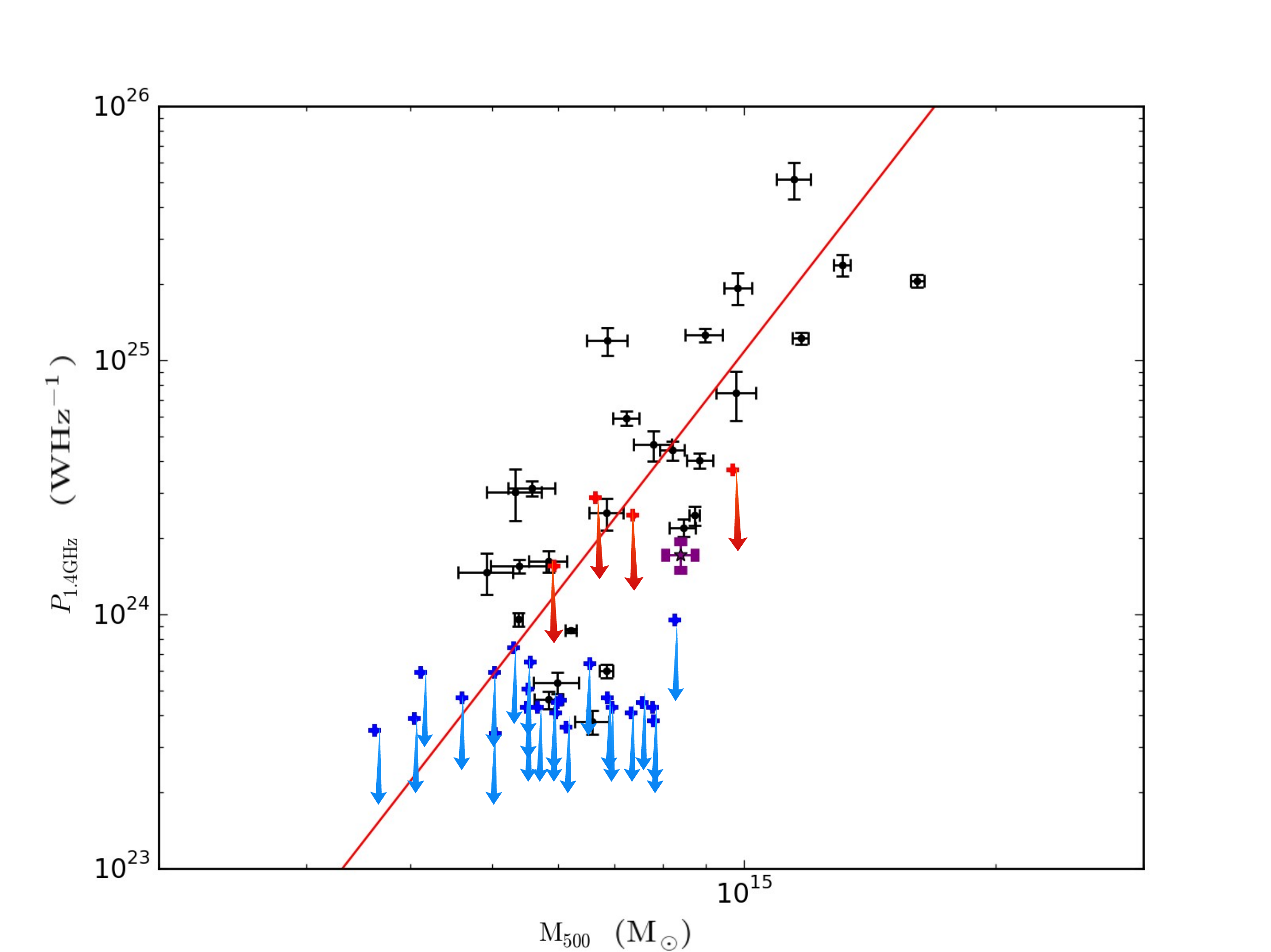}
      \caption{Synchrotron power of RHs at 1.4 GHz ($P_{1.4\,GHz}$) vs. cluster mass M$_{500}$ \citep[from][]{2016A&A...594A..27P}. Detected radio halos reported in the literature measured at 1.4 GHz \citep[see][]{2016A&A...595A.116M} are shown in black dots. The red line is the best fit for the black dots. The purple star corresponds to the radio halo on PSZ2 G284.97-23.69. The blue arrows correspond to detection limits appearing in \cite{2013ApJ...777..141C} and references therein. The red arrows are the upper limits for the non-detections presented in this paper (see Table \ref{table:upperlimits}).
             }
         \label{fig:Fighalos}
\end{figure*}

In Fig. \ref{fig:Fighalos} we show the upper limits for the non-detections in the ATCA sample as red arrows on the $P_{1.4\,GHz}$ vs. M$_{500}$ plot. We compare our calculated values with those obtained by \citet{2008Venturi} and \citet{2013A&A...557A..99K} (blue arrows in Fig. \ref{fig:Fighalos}), measured using GMRT images, with the modelled halo injections performed at 610 MHz. Our calculated ATCA upper limits appear higher in the plot as a consequence of the combination of two effects. The first is that the redshift range of our sample is higher than previous studies (effectively limited to z < 0.3, see Sect. \ref{sect:INTRO}). In our calculation of the radio power, we have the luminosity distance factor that scales as  $D_{\rm Lum}(z)^2$. The second and most important effect is the frequencies of the ATCA observations. Our upper limit injections were performed at $\sim$ 1.9 GHz, and assuming a typical spectral behaviour of RHs (with spectral index $\alpha$=1.3), we obtained higher values of the radio power when we rescale to 1.4 GHz. The values of our calculated upper limits appear in Table \ref{table:upperlimits}. It is worth mentioning that the power of the giant RH discovered in PSZ2 G284.97-23.69 and some other RHs reported in the literature lie close to (or even below) the radio power regime of the upper limits obtained for our ATCA observations. This could imply that the measured flux density of our detected sources could be underestimated.

\begin{table}[h]
            \caption[]{Upper limits obtained for the ATCA cluster images without traces of diffuse radio emission. The calculated logarithm of the radio power at 1.4 GHz, by assuming a spectral index of 1.3, for the total flux density of the injected fake radio halo model in the Block 3 sub-band (1.9 GHz) appears in Column 2.}
\label{table:upperlimits}
$$\begin{tabular}{c   c }
\hline
\hline
Cluster name &      Log $P_{1.4GHz}$  \\
         &          (WHz$^{-1}$)         \\
\hline
PSZ2 G285.63-17.23     &     24.46  \\
PSZ2 G277.76-51.74     &     24.57  \\
PSZ2 G286.28-38.36     &     24.19  \\
PSZ2 G271.18-30.95     &     24.39  \\
\hline
\hline
\end{tabular}$$ 
\end{table}

\section{Discussion and conclusions}\label{sect:conclusions}

A full analysis in search of diffuse radio emission on the ATCA sub-sample of the MACS-Planck RHCP is presented in this paper.

Among the seven targets, only two clusters show a confirmed or tentative diffuse radio emission. We reported the presence of a giant radio halo in PSZ2 G284.97-23.69 and found a new candidate diffuse source in PSZ2 G262.73-40.92. The X-ray morphological analysis based on \xmm\ observations indicates that those objects are most likely disturbed systems, based in particular on the centroid shift indicator ($w$), with a concentration parameter ($C$) not far from the threshold adopted to separate merging and relaxed clusters. 

This is not totally unexpected. Our choice for the limits for cool-core/non cool-core and relaxed/disturbed systems is based on the limits for $C$ and $w$, which were adopted and obtained empirically from the local (z = 0.055 - 0.183) \rexcess\ cluster sample \citep{2007A&A...469..363B}, in which two disturbed cool-core clusters are detected \citep{2009A&A...498..361P, 2010A&A...514A..32B}. Furthermore, the results of \citet{2017arXiv170205094M} on the SPT sample suggest that the evolution of the core and outer regions are decoupled, which would indicate that there is no reason why a cluster cannot be a cool core and disturbed as well.

\begin{figure}[!h]
   \includegraphics[width=10cm]{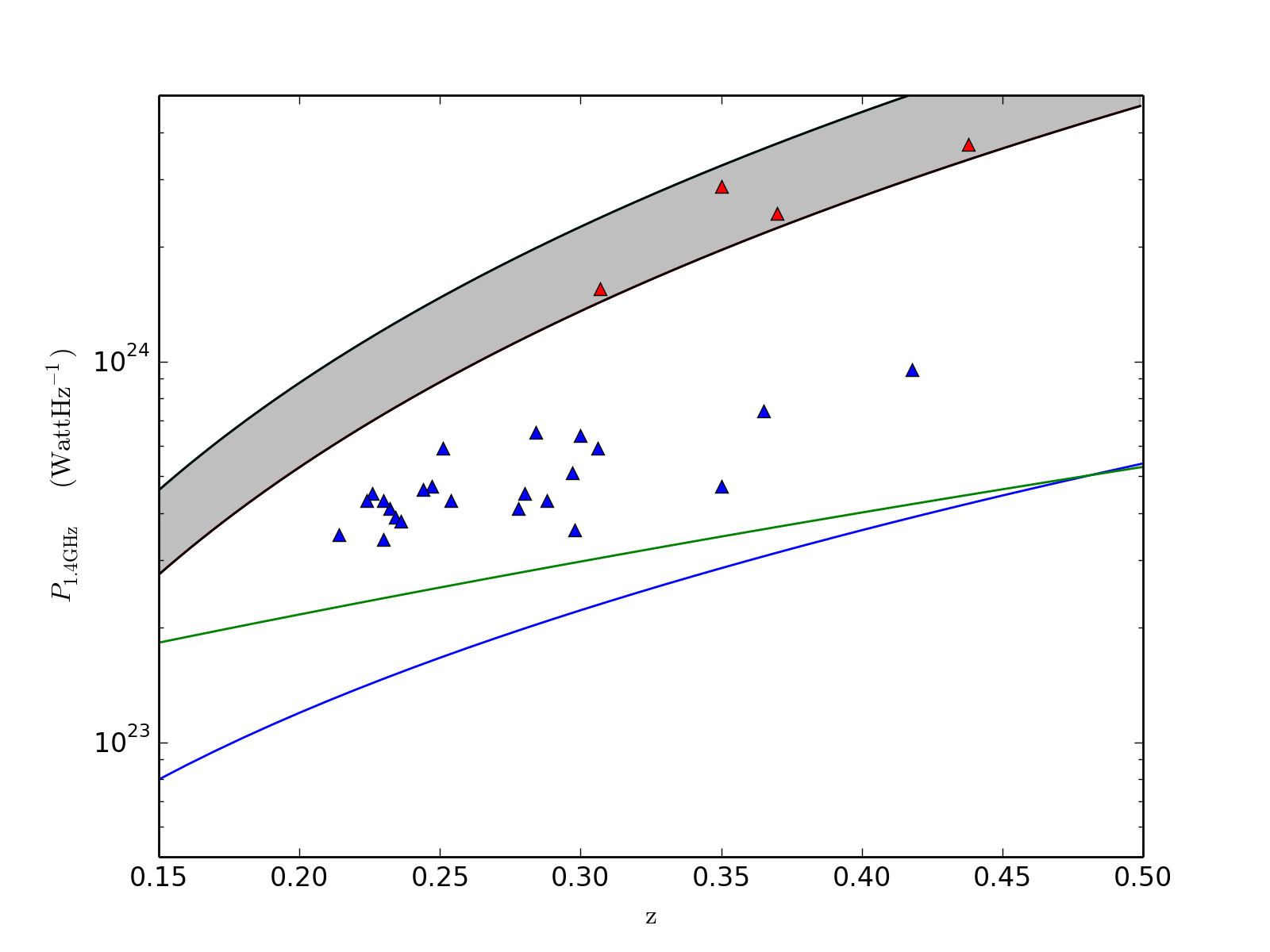}
      \caption{Upper limit radio powers rescaled at 1.4 GHz vs. redshift for our ATCA observations (red triangles) and GMRT observations Venturi (2008) \& Kale (2013) (blue triangles). The lower (upper) border of the grey band corresponds to the lowest (highest) detection limit of injected fluxes, i.e. 3 mJy (5 mJy) for our ATCA images. The green and blue lines show the minimum power at 1.4 GHz of detectable radio halos in the EMU survey as derived  in \citet{2012A&A...548A.100C}. } 
        \label{fig:limitsredshift}
\end{figure}
 
Those clusters of our sample that have clear X-ray morphological indicators that they are highly disturbed do not present any bright diffuse emission. Currently, these clusters (upper limits) are still consistent with being on the $P_{1.4}-M_{500}$ correlation. Future deeper observations will have to determine if these halos could be under-luminous and fall below the correlation. Hypothetical RH could be either ``classical'' halos, i.e. almost consistent with the correlation, or under-luminous RHs with ultra-steep radio spectra, detectable with low frequency radio observations \citep[these are particularly expected in high-z clusters; see][]{2006Cassano}. The challenging hypothesis to be tested would be if those clusters are radio quiet. There are actually cases of disturbed clusters lacking radio halos. \citet{2011Russell} have shown that the galaxy cluster A2146 possesses no diffuse emission even though it is clearly a merging cluster \citep[see also][for a detailed study of the dynamical status of the cluster]{2015White}. However, in this particular case the mass of the cluster is very low (4.3 $\times$ 10$^{14}$ M$_{\odot}$), which could be the reason for the lack of a detectable radio halo. 

Indeed, upper limits indicated in Fig. \ref{fig:Fighalos} are significantly higher compared to other works \citep{2013ApJ...777..141C}, even if the rms sensitivities reached in all these analyses are comparable. This is explained by the fact that previously published upper limits are based either on lower frequency (mostly 610 MHz) observations \citep{2008Venturi, 2013A&A...557A..99K} or on lower redshift ATCA cluster samples \citep{2016MNRAS.459.2525S}. In the first case (and keeping in mind that we convert radio powers to the reference frequency of 1.4 GHz), the same rms sensitivity at 1.9 GHz and 610 MHz translates to a higher 1.4 GHz radio halo luminosity upper limit for the 1.9 GHz case because of the spectral steepness of radio halos \citep[we rescaled by assuming a spectral index $\alpha$=1.3 to be consistent with][]{2013ApJ...777..141C}. In the second case, the clusters are observed at the same frequency, but located at z$\sim$0.15, which, owing to the $D_{\rm Lum}(z)^2$ factor, translates into $\sim$10 higher upper limits in radio power compared to lower redshift ATCA observations. 

To be able to distinguish between radio-quiet or under-luminous diffuse emission in the three disturbed galaxy clusters in our sample, it is clear that more sensitive and lower frequency radio observations are needed. Instruments such as the JVLA, the upgraded GMRT, or the LOFAR  Two-metre Sky Survey \citep[see e.g.][]{2017A&A...598A.104S} are already improving the observational capabilities needed for such studies by reaching better sensitivities. As an example of the expected power of future instruments, in Fig. \ref{fig:limitsredshift} we compare the detection upper limits obtained for our ATCA observations (red triangles) with previous upper limits taken with the GMRT at 610 MHz (blue triangles) as a function of redshift. The green and blue lines show the minimum power at 1.4 GHz of detectable radio halos in the Evolutionary Map of the Universe (EMU) survey for different signal-to-noise levels as derived  in \citet{2012A&A...548A.100C} (see Sect. 5 and Eqs. 9 and 10 in their paper). The EMU survey will produce a deep radio continuum survey of the southern sky with rms noises of $\sim$ 10$\mu$Jy/beam; these levels are comparable to our best rms noise of $\sim$11$\mu$Jy/beam, but have lower resolution \citep[$\sim$ 10 arcseconds;][]{2011PASA...28..215N}, thereby resulting in a better sensitivity at lower frequencies (1.3 GHz). From Fig. \ref{fig:limitsredshift} we note that EMU will have the sensitivity to discover possible halos in clusters that presently are undetected in both GMRT and ATCA samples.

\begin{acknowledgements}
The authors would like to warmly thank Roberto Ricci for useful discussions during the preparation of this paper. GMA is supported by the Erasmus Mundus Joint Doctorate Program by Grants Number 2013-1471 from the agency EACEA of the European Commission. GMA, CF, and MJ-H acknowledge financial support from {\it Programme National Cosmologie et Galaxies (PNCG)} and {\it Universit\'e de Nice-Sophia Antipolis -- Programme Professeurs Invit\'es 2015}. MJ-H acknowledges the Marsden Fund administered by the Royal Society of New Zealand on behalf of the Ministry of Business, Innovation, and Employment. TV, RC, and GB acknowledge partial support from PRIN-INAF 2014 grant. The research leading to these results has received funding from the European Research Council under the European Union’s Seventh Framework Programme (FP7/2007-2013) / ERC grant agreement no. 340519. Basic research in radio astronomy at the Naval Research Laboratory is supported by 6.1 Base funding. The Australia Telescope Compact Array is part of the Australia Telescope National Facility, which is funded by the Australian Government for operation as a National Facility managed by CSIRO. The results of these paper are partially based on data retrieved from the ESA Planck Legacy Archive, NASA SkyView, and the Hubble Legacy Archive, which is a collaboration between the Space Telescope Science Institute (STScI/NASA), the Space Telescope European Coordinating Facility (ST-ECF/ESA), and the Canadian Astronomy Data Centre (CADC/NRC/CSA). 
\end{acknowledgements}


\bibliographystyle{aa}
\bibliography{plck285}

\begin{appendix}
\section{High resolution radio images of ATCA sample}\label{sect:appendix}

   \begin{figure*}[ht]
   \centering
   \includegraphics[width=8cm]{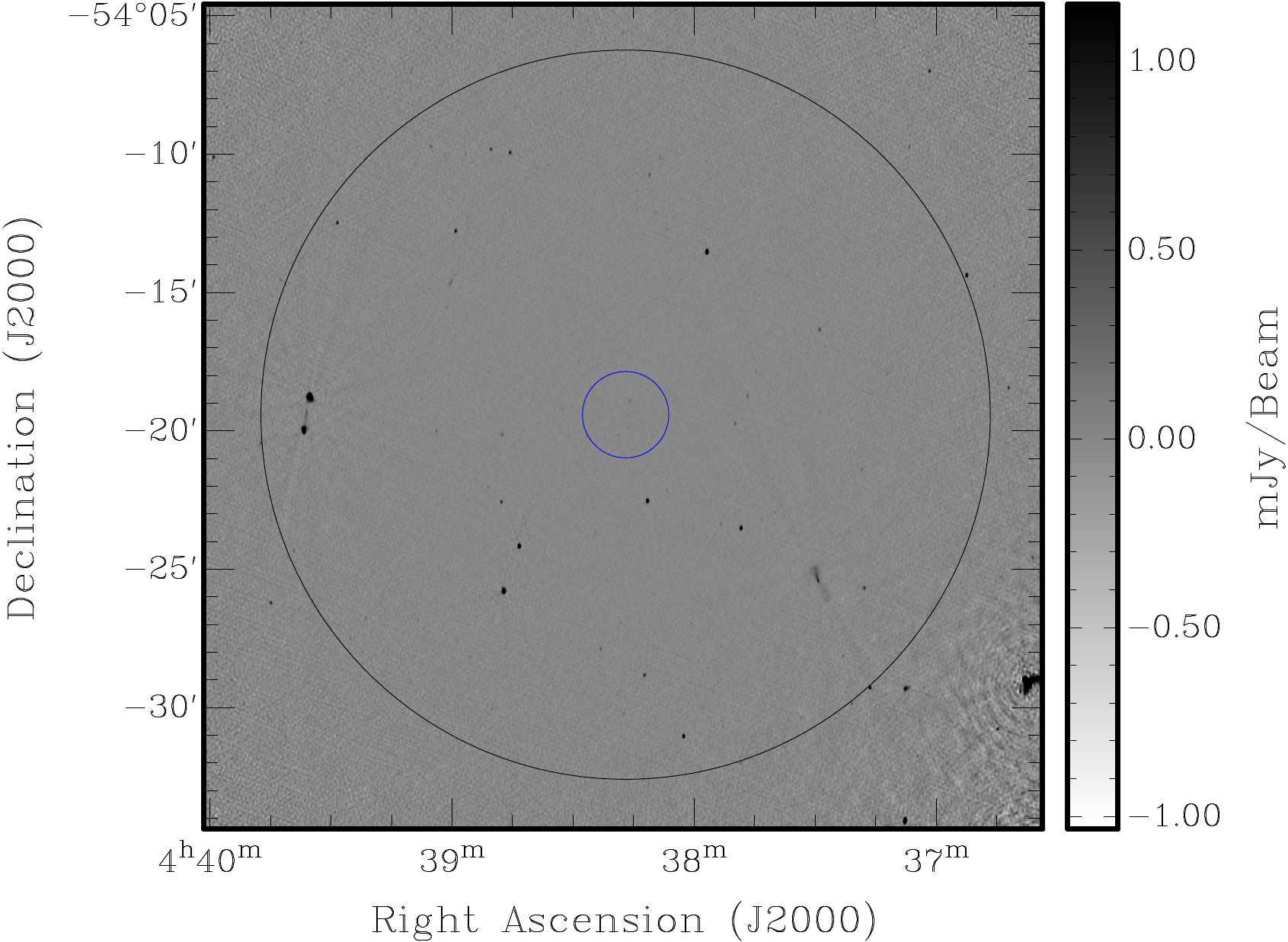}
   \includegraphics[width=8cm]{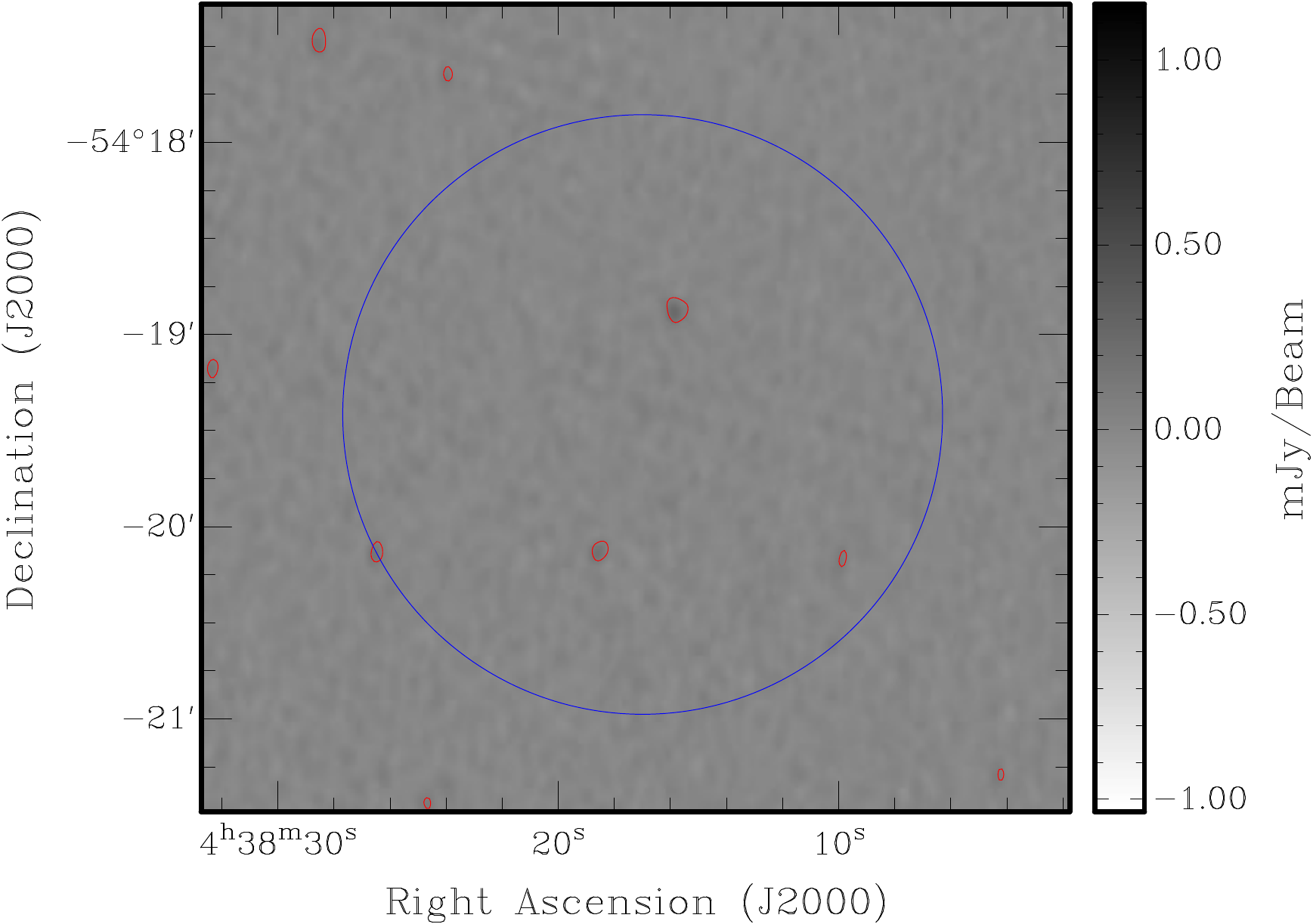}
   \includegraphics[width=8cm]{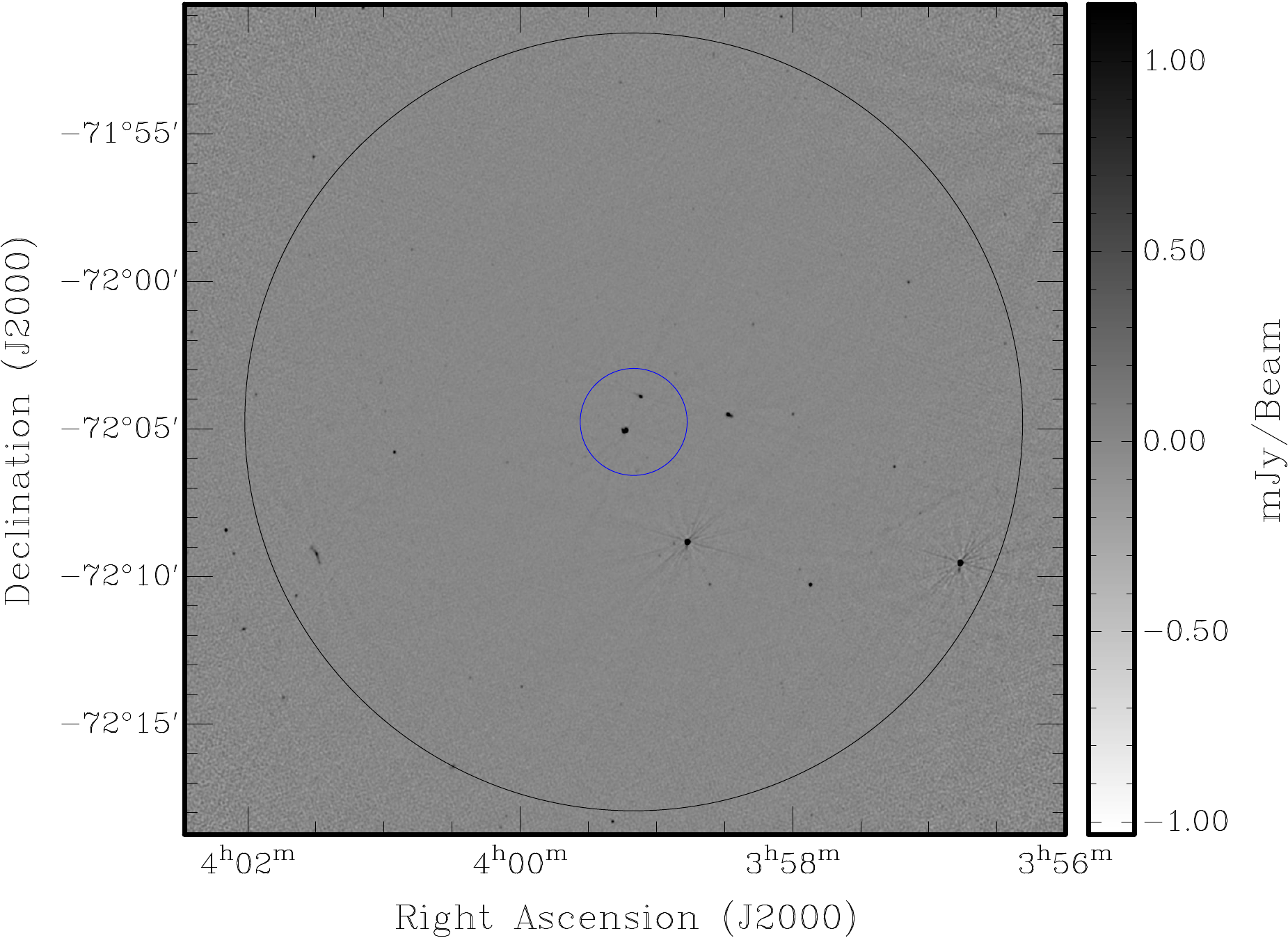}
   \includegraphics[width=8cm]{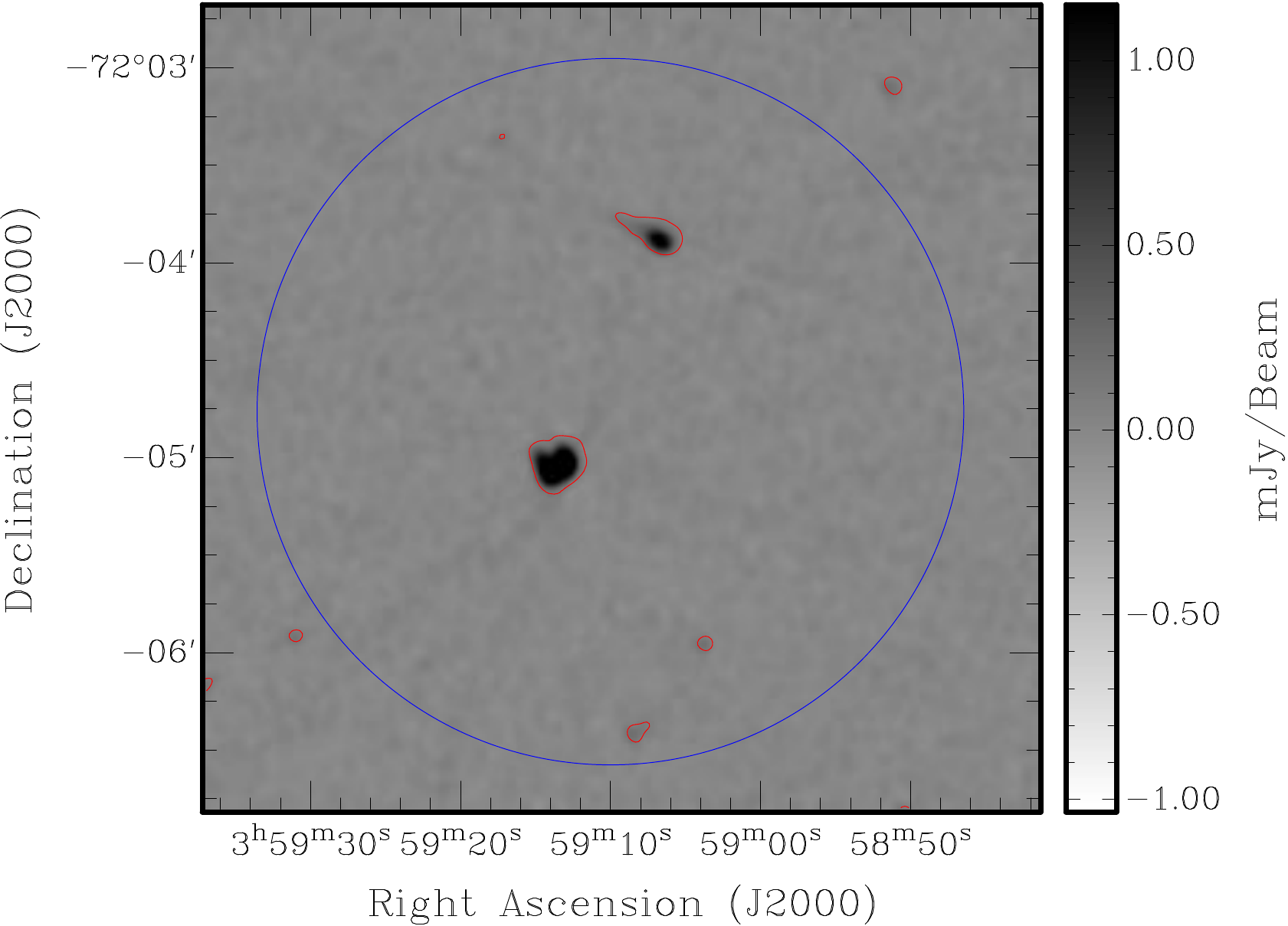}
   \includegraphics[width=8cm]{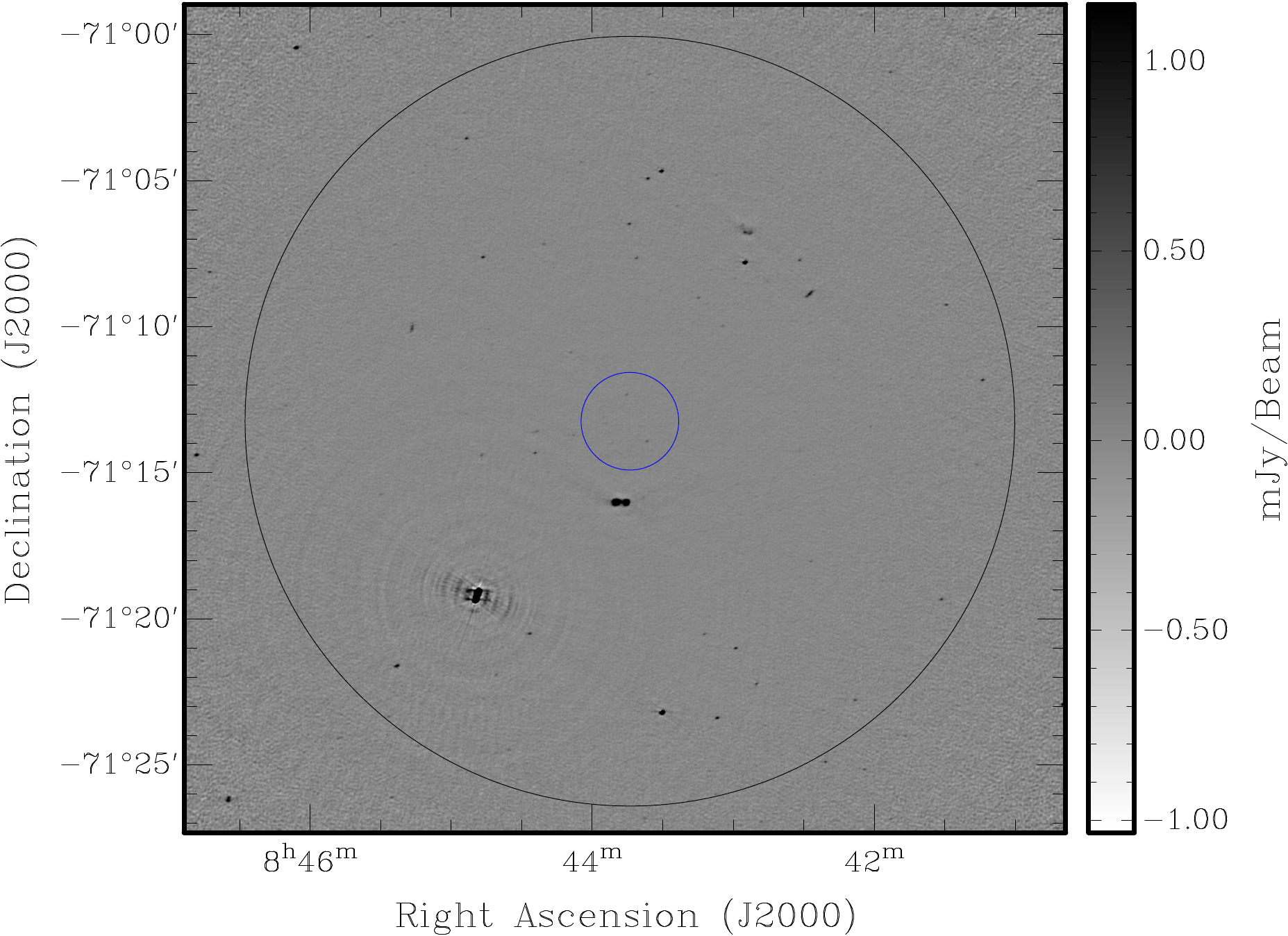}
   \includegraphics[width=8.3cm]{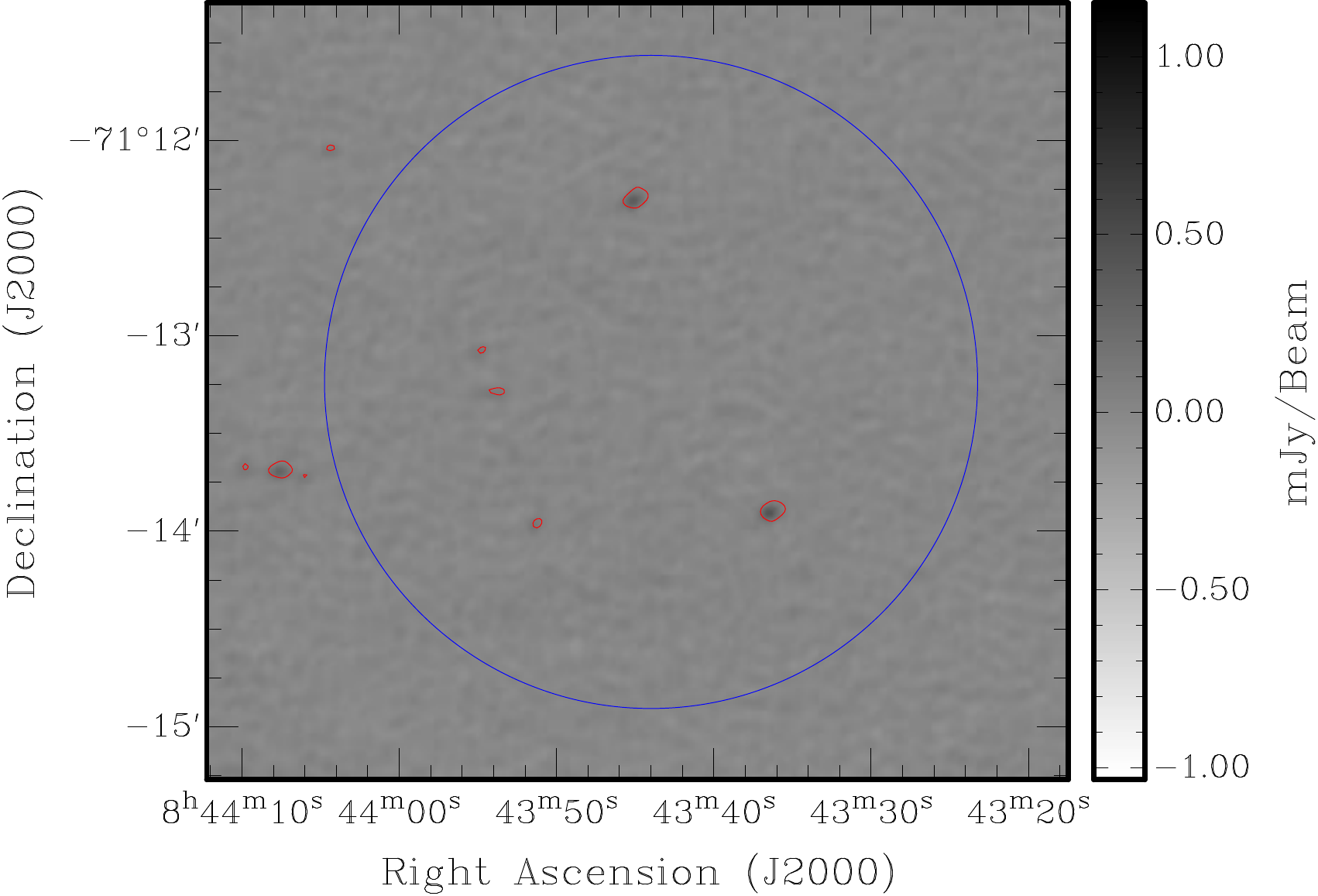}
      \caption{Final full-resolution, wide-band ATCA images of the cluster sample. \textit{Left panels}: Full field image centred on cluster coordinates. The outer circle denotes the boundary of the primary beam with a radius of $\sim$0.22 degrees. The central 1 Mpc-diameter region is indicated by the inner, blue smaller circle. \textit{Right panels}: Zoom into the central area with 3$\sigma$ contours of the corresponding map overlaid in red. \textit{Top}: PSZ2 G262.73-40.92; \textit{Middle}: PSZ2 G286.28-38.36; \textit{Bottom}: PSZ2 G285.63-17.23.
             }
         \label{fig:Figwide1}
   \end{figure*}

   \begin{figure*}[ht]
   \centering
   \includegraphics[width=8.5cm]{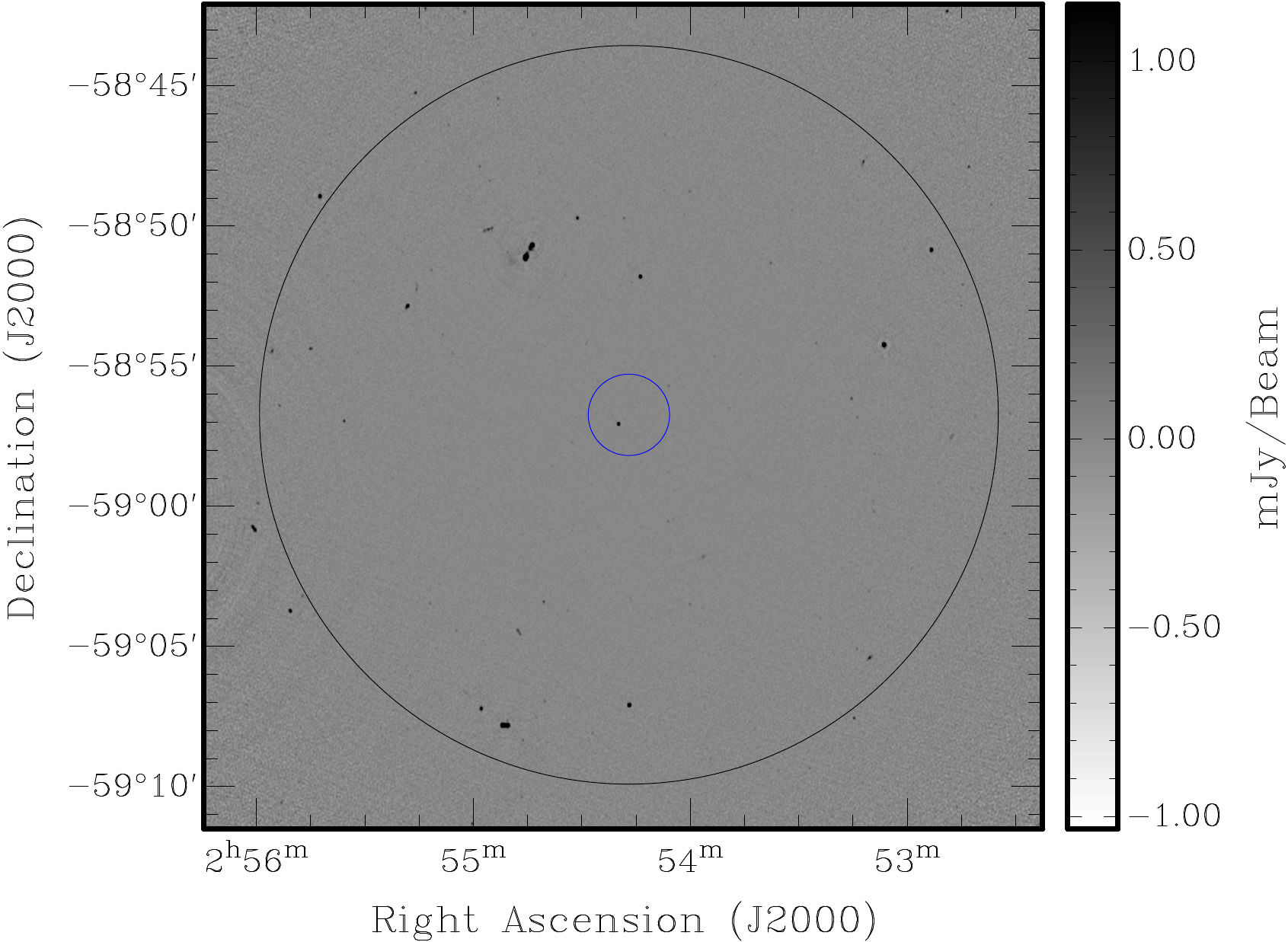}
   \includegraphics[width=9cm]{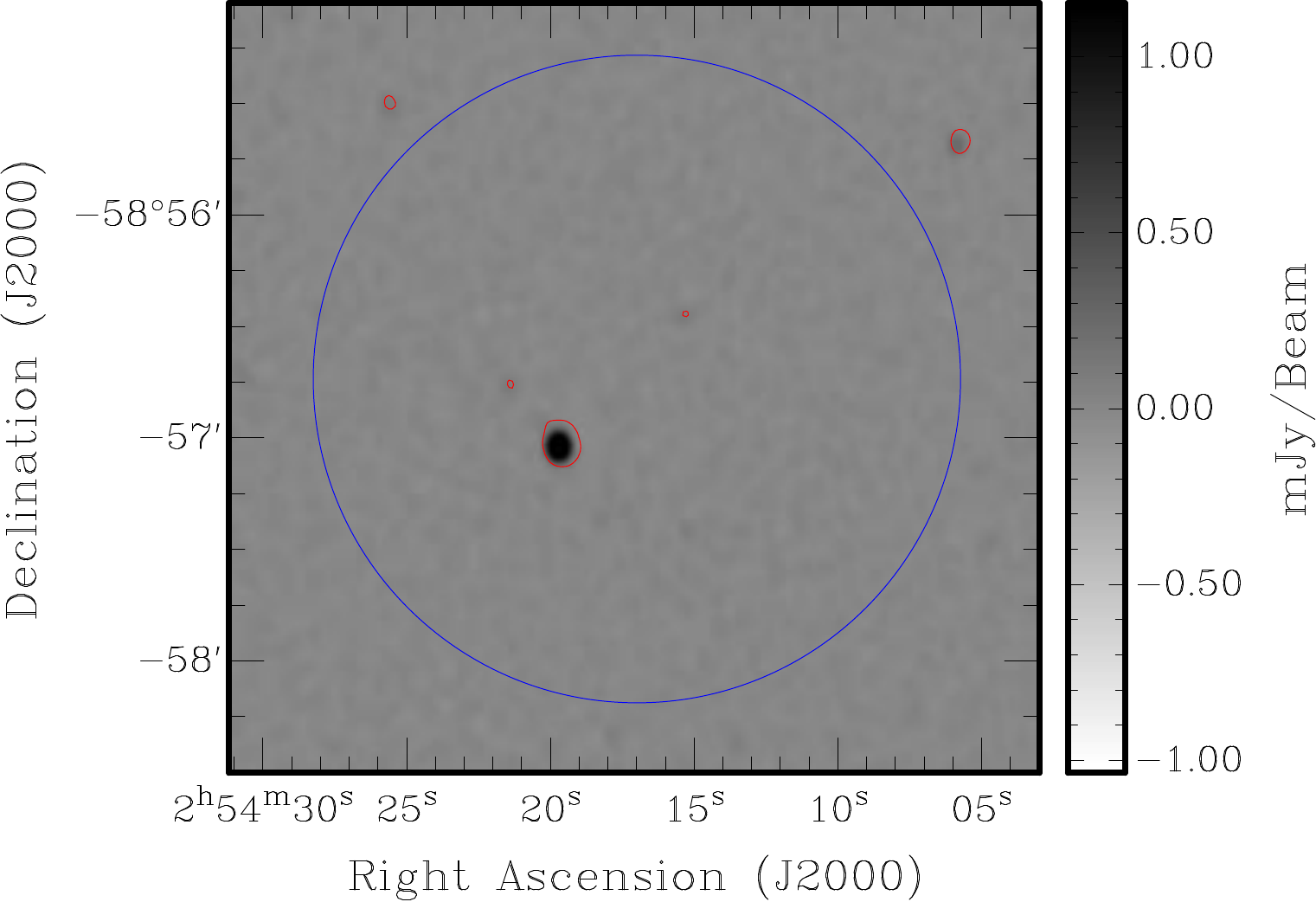}
   \includegraphics[width=8.3cm]{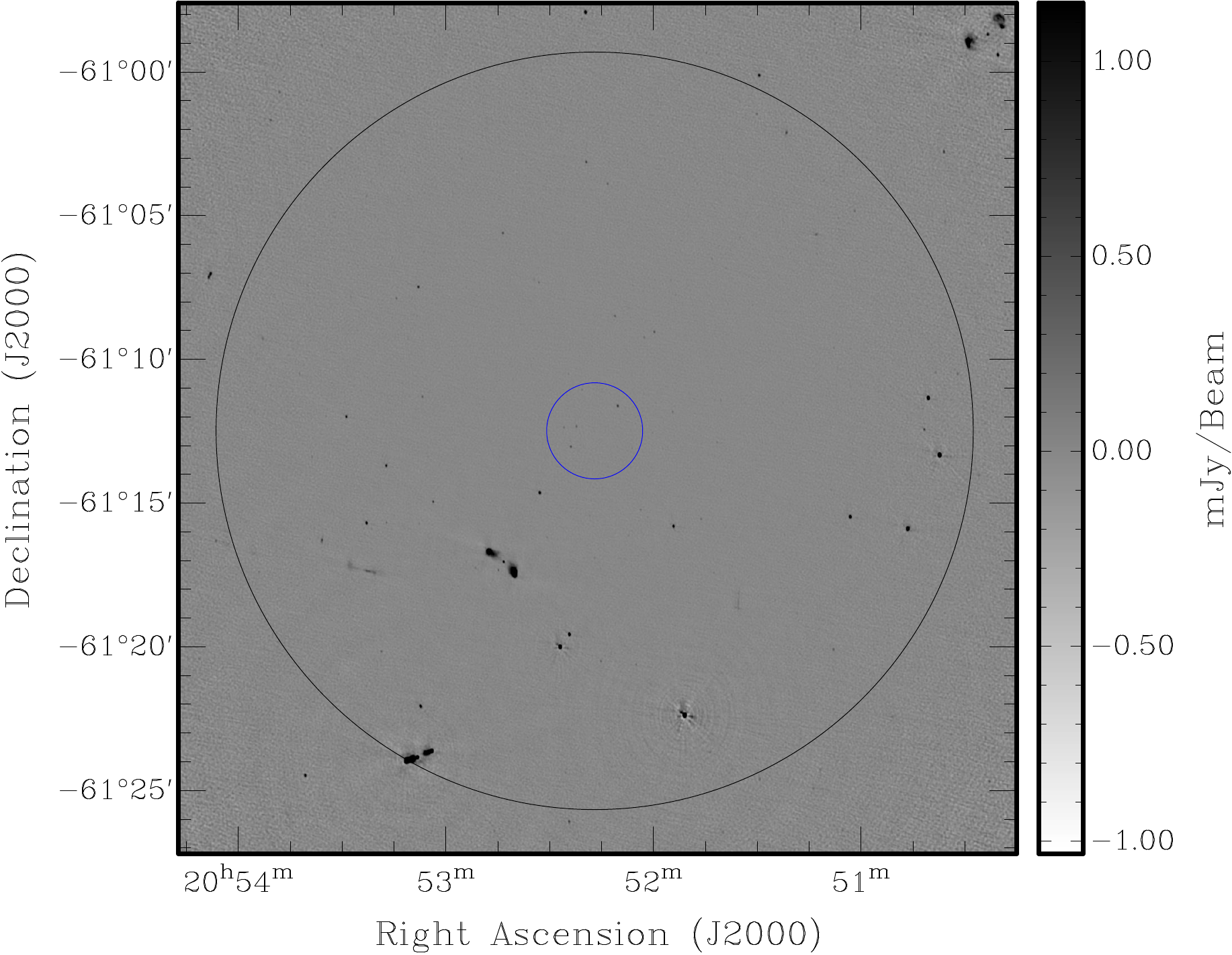}
   \includegraphics[width=9cm]{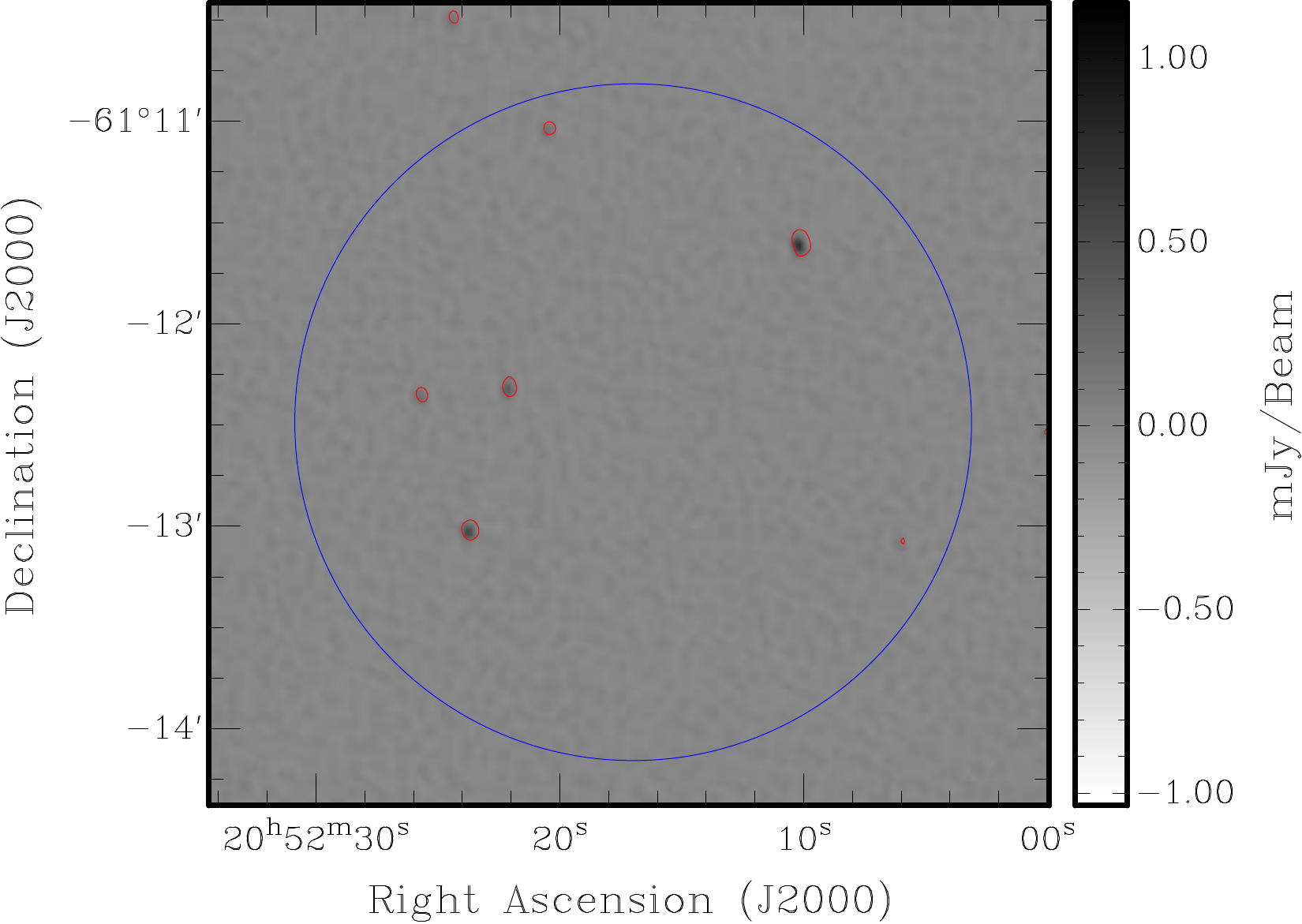}
   \includegraphics[width=9cm]{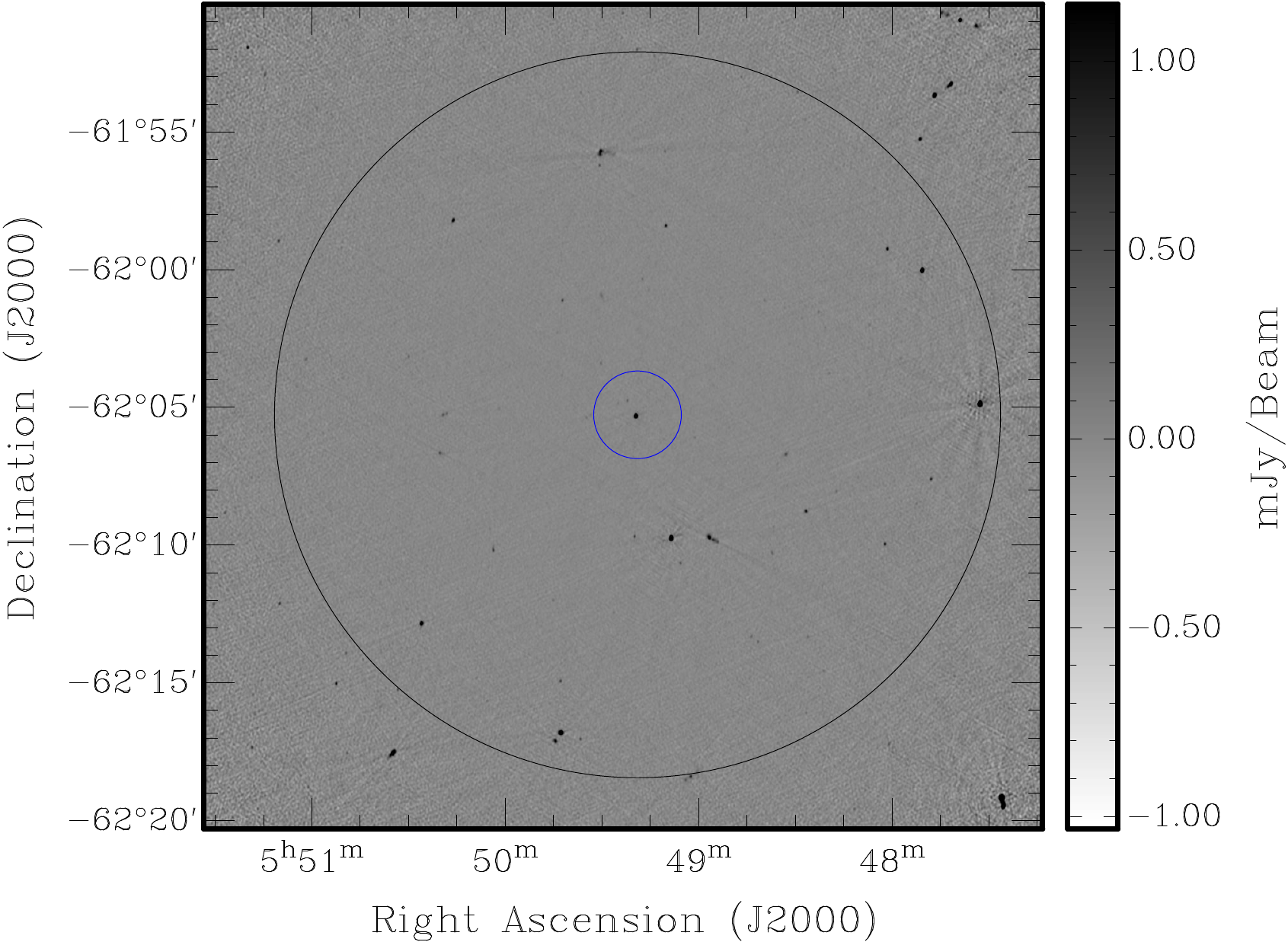}
   \includegraphics[width=9cm]{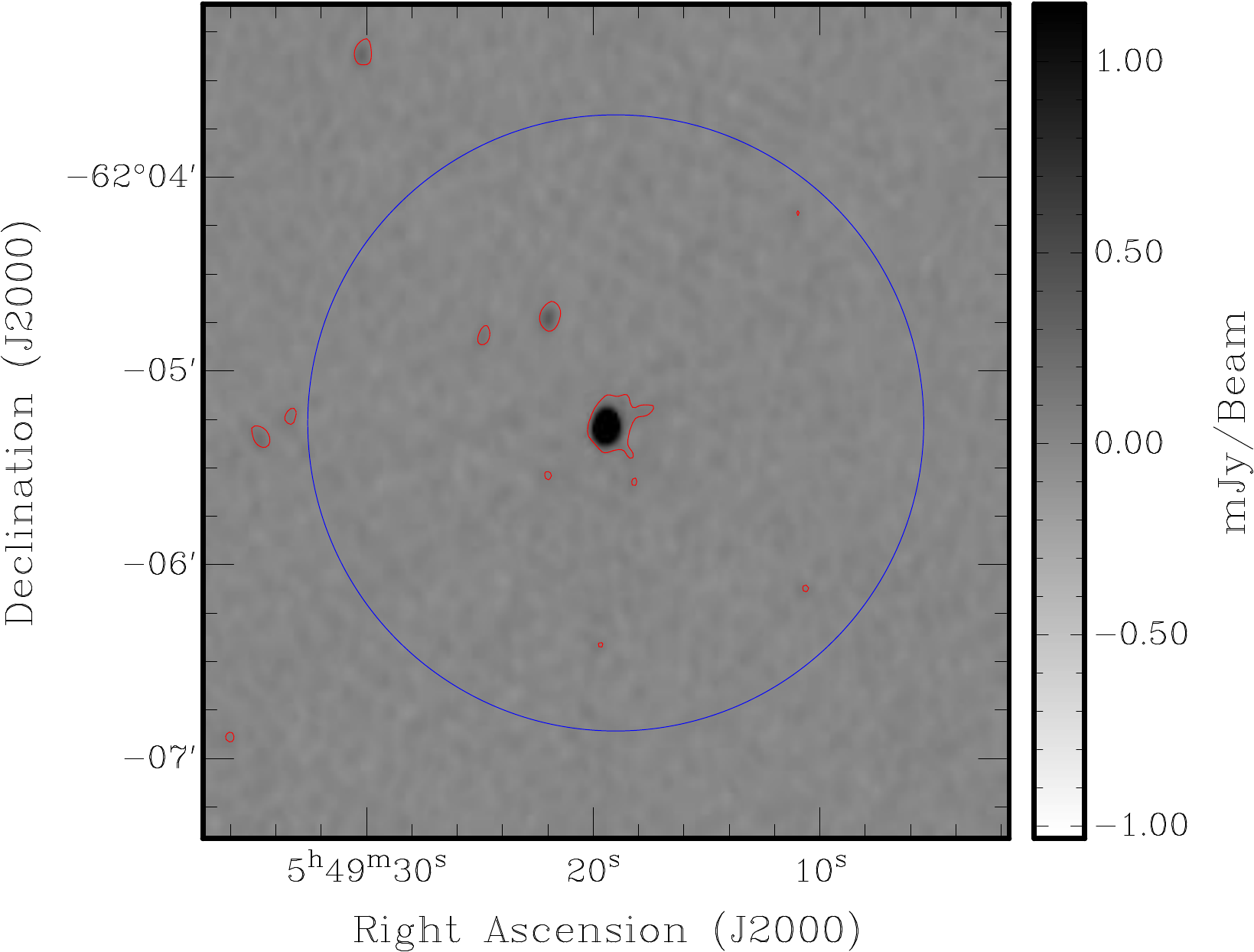}        
      \caption{Same as Figure \ref{fig:Figwide1}. \textit{Top}: PSZ2 G277.76-51.74; \textit{Middle}: PLCK G334.8-38.0; \textit{Bottom}: PSZ2 G271.18-30.95.
             }
         \label{fig:Figwide2}
   \end{figure*}

\end{appendix}

\end{document}